\documentclass{article}
\usepackage{authblk}

\usepackage{arxiv}
\usepackage{natbib}
\usepackage{float}      % 支持 [H] 定位选项
\usepackage[utf8]{inputenc} % allow utf-8 input
\usepackage[T1]{fontenc}    % use 8-bit T1 fonts
\usepackage{hyperref}       % hyperlinks
\usepackage{url}            % simple URL typesetting
\usepackage{booktabs}       % professional-quality tables
\usepackage{amsfonts}       % blackboard math symbols
\usepackage{nicefrac}       % compact symbols for 1/2, etc.
\usepackage{microtype}      % microtypography
\usepackage{lipsum}
\usepackage{graphicx}
\graphicspath{ {./images/} }

\usepackage{hyperref}
\usepackage[cmex10]{amsmath}
\usepackage{amssymb}
\usepackage{multirow}

\newtheorem{proposition}{Proposition}

\newtheorem{definition}{Definition}
\newtheorem{example}{Example}

\newtheorem{remark}{Remark}

\title{Optimization Models to Meet the Conditions of Order Preservation in the Analytic Hierarchy Process}

\author[1,2]{Jiancheng Tu}
\author[2]{Wu Zhibin}
\author[2]{Yueyuan LI}
\author[2,3]{Chuankai XIang}

\affil[1]{Department of Computing, Hong Kong Polytechnic University, Hong Kong}
\affil[2]{Business School, Sichuan University, Chengdu 610065, China}
\affil[3]{ City University of Hong Kong, School of Data Science, Kowloon Tong, Hong Kong}

\begin{document}
\maketitle
\begin{abstract}
Deriving a priority vector from a pairwise comparison matrix (PCM) is a crucial step in the Analytical Hierarchy Process (AHP). 
Although there  exists  a priority vector that satisfies  the conditions of order preservation (COP), the priority vectors obtained through existing prioritization methods frequently violate these conditions, resulting in numerous COP violations.
To address this issue, this paper introduces a novel procedure to manage COP violations in AHP. 
Firstly, we prove that the index-exchangeability condition is both a necessary and sufficient condition for determining whether a priority vector satisfies COP. 
This  enables the direct detection of COP violations, relying solely on the pairwise comparison preferences of decision-makers, rather than the prioritization methods utilized.
Subsequently, we propose the Minimal Number of Violations and Deviations Method (MNVDM) model, which aims to derive a priority vector with the minimal number of COP violations. In particular, the MNVDM can obtain a violation-free priority vector when the PCM meets the index exchangeability conditions. 
Furthermore,   an optimization model  based on minimizing information loss is designed to  ensure the COP by revising the preferences when the index-exchangeability conditions are violated.
Finally, the feasibility and efficiency of the proposed models are validated through numerical examples and Monte Carlo simulation experiments.
Our implementation is  available at: 
\href{https://github.com/Tommytutu/COP}{https://github.com/Tommytutu/COP}.
\end{abstract}

% keywords can be removed
\keywords{ Multiple-criteria decision analysis \and analytical hierarchy process \and pairwise comparison matrix \and  conditions of order preservation
}

\section{Introduction}
\label{sec:introduction}
The Analytical Hierarchy Process (AHP), proposed by Saaty \cite{Saaty1980}, is one of the most widely utilized  multi-criteria decision-making methods. 
AHP encompasses five primary steps: structuring the hierarchy, constructing pairwise comparison matrices (PCMs), checking consistency, deriving the priority vector, and synthesizing the results. 
This paper focuses  specifically on the procedures of checking consistency and deriving the priority vector.

AHP distinguishes itself from other methods through its use of pairwise comparison techniques to establish the weights of criteria at the same level \cite{brunelli2015}.
This method's effectiveness lies in its ability to break down complex decision-making problems into simpler, more manageable sub-problems, thereby allowing decision-makers (DMs) to systematically compare the relative importance of each criterion.
However, the subjective nature of judgments can introduce inconsistencies \cite{Brunelli2020b}. 
Scholars have thus proposed diverse measures to assess the rationality of DMs' judgments, ranging from consistency indices \cite{brunelli2018, grzybowski2016new,kazibudzki2023,brunelli2024}, transitivity (also known as ordinal consistency) \citep{WuZhibin2021Mtac,brunelli2020,Xu2021,Liu2021}, to conditions of order preservation (COP) \cite{BanaECostaCarlosA2008,kulakowski2015,kulakowski2019,cavallo2020,mazurek2022}.

The consistency ratio (CR) \cite{Saaty1980} is the most widely  used consistency index.
Saaty \cite{Saaty1980} suggested that the inconsistency level of  the PCM  is acceptable when its CR value is below 0.1.
In cases where the CR exceeds this threshold, the DM is advised to revisit their preferences to ensure greater consistency.
However, Saaty's concept of consistency measurement has been seriously questioned \citep{genest1994,karapetrovic1999,koczkodaj2018}.
As a result, numerous other inconsistency indices have been developed from various perspectives.
These include the minimal error index (MEI) \citep{kulakowski2015,kulakowski2019}, the geometric consistency index (GCI) \cite{Aguarón2003}, the index of logarithmic squared deviations method (ILSDM) \citep{kazibudzki2022,kazibudzki2021}. For further exploration of inconsistency indices, interested readers can refer to recent surveys on the topic  \cite{brunelli2018,pant2022}.

Deriving the priority vector from the PCM is a crucial step in AHP after conducting pairwise comparisons.
The  eigenvalue method (EM), as proposed by Saaty \cite{Saaty1980}, is the most popular prioritization method.
However, EM has faced criticism from academics due to fundamental weaknesses when dealing with inconsistent PCMs. 
For example, it has been noted that the weight is not invariant under transpositions\citep{Johnson1979},   is inefficient \citep{bozoki2018}, and violates the monotonicity axioms \citep{csato2021}.
Therefore, various alternative prioritization methods have been proposed \citep{CHOO2004}.
Many of these  aim to minimize the deviations between the estimated consistent PCM and the original PCM, sharing a common framework\citep{CHOO2004,kazibudzki2019q}:

\begin{equation}\label{DMPM}
\text { min } {\cal{D}}\left(\mathbf{A}, \left[w_{i} / w_{j}\right]\right)
\end{equation}
with some accompanying constraints such as positive coefficients and the normalization conditions. 
When $\mathbf{A}$ is inconsistent, different definitions of  function ${\cal{D}}\left(\mathbf{A}, \left[w_{i} / w_{j}\right]\right)$ can yield varying results. 
Besides, the objective of model (\ref{DMPM}) can be interpreted as the inconsistency index.
Specially, the logarithmic least squares method (LLSM) put forth by Crawford and Williams \citep{Crawford1985}  has gained increasing prominence owing to its sound axiomatic justifications \citep{csato2019,csato2021,bozoki2018}.
Undoubtedly, several important work on prioritization methods has been reported in recent years \cite{kulakowski2015,kazibudzki2022,kazibudzki2021,Zhang2021,Faramondi2020,yuen2024}.

As per  Saaty \cite{saaty2003}, the priority vector of a set of alternatives possesses two distinct interpretations, stating that
``The first is a numerical ranking of the alternatives that indicates an order of preference among them. The other is that the orders should also reflect intensity or cardinal preference as indicated by the ratios of the numerical values and is thus unique to within a positive multiplicative constant (a similarity transformation)."
In other words, the priority vector should contain information regarding the order and intensity of the preferences.
Bana e Costa and Vansnick \cite{BanaECostaCarlosA2008} concurred with this and articulated these two meanings of priority vectors as the COP, which encompass the preservation of order preference  (POP) conditions  and the preservation of order of intensity preference (POIP) conditions.
They emphasized that when the PCM is inconsistent, the priority vector obtained by the EM and LLSM sometimes cannot appropriately reflect the DM's preferences. 
That is, the EM and LLSM  may fail to generate a priority vector from the acceptable PCM ($\text{CR}\leq 0.1$)  meeting the COP,
and the consistency measures (such as CR and GCI) may not be capable of detecting or warning of this `violation' phenomenon  (see Example \ref{eample11}).

The ranking outcomes are deemed indisputable when the priority vector adheres to the COP. 
Conversely, if the priority vector violates COP, the DM may resist accepting ranking results that appear unreasonable. 
Complying with COP can thus reduce the costs associated with implementing the evaluation system, offering advantages such as providing the ranking participants with a sense of fairness and transparency.

Therefore, this paper develops optimization models to ensure adherence to the COP within the AHP.
The subsequent sections of this paper are structured  as follows:
Section \ref{22}  provides a review of relevant literature concerning COP.
Section \ref{sec3} introduces some fundamental concepts. 
Section \ref{sec4}   outlines a methodology for identifying COP violations.
 Section \ref{sec5}  introduces the minimal number of violations and  deviations method (MNVDM)    for deriving priority vectors from  inconsistent PCMs.
Section \ref{sec6} designs an optimization model  to  revise the  preferences when the violations are  inevitable.
The conclusions are given in Section \ref{sec7}.

\section{Related work}\label{22}
This section first reviews the related work about the conditions of order preservation (COP), and then discusses the  motivations and contributions of this paper.

For convenience, let $N=\{1,2,3,...,n\}$, and  let  ${X}=\{x_1,x_2,\ldots,x_n\}$  be the given set of alternatives.
$\mathbf{A}=(a_{ij})_{n\times n} \subset X \times X$ is a  pairwise comparisons matrix (PCM), which satisfies the  reciprocality, that is, $a_{ij}\cdot a_{ji}=1, i,j\in{N}$.
$ \mathbf{w}= {({w_1},{w_2},\ldots,{w_n})^T} $ satisfying $\sum\limits_{i= 1}^n w_i=1$ and $w_i > 0$ is the priority vector derived from $\mathbf{A}$.
\subsection{Conditions of order preservation}
Regarding the issue of COP violations, two main questions need to be addressed: the underlying causes of COP violations and the potential strategies for mitigating these violations.
For the first question, some researchers argue that COP violations are not intrinsic to PCM itself but are a consequence of inconsistencies \citep{kulakowski2015,kulakowski2019}.
Mazurek and Ku{\l}akowski \cite{JiriMazurek2020} examined the satisfaction of the COP across various levels of inconsistency using the Monte Carlo simulation method.
Their findings indicate that, for PCMs with lower inconsistency, both the EM and the LLSM maintained COP almost identically.
However, these studies fail to consider inherent flaws in the prioritization methods that might also contribute to COP violations.
As shown in Example \ref{eample11}, although there is a priority vector that satisfies COP,   priority vectors by EM and LLSM violate COP.
Therefore, we propose that the causes of COP violations include both inconsistency and flaws in the prioritization methods.
Consequently, it is imperative to develop a prioritization method that accounts for COP.

To address COP violations, several significant studies have been conducted \citep{siraj2012,cavallo2019,cavallo2020,Brunelli2020b,mazurek2022,Tuwu2022}.
Siraj \cite{siraj2012} designed an optimization model aimed at determining the priority vector by minimizing the number of POP violations.
Similarly, Tu and Wu \cite{Tuwu2022} devised   a mixed-integer linear optimization model to concurrently satisfy the POP conditions and achieve an acceptable level of inconsistency by adjusting some inconsistent preferences.
However, given that POIP is a more stringent criterion than POP (see Definition \ref{defPOIP}), these studies focusing on POP  \citep{siraj2012,Faramondi2020,Tuwu2022,tu2023} are not applicable to POIP. 

Cavallo and D'Apuzzo \cite{cavallo2020} found  a necessary condition, termed  the index-exchangeability condition,  which must be satisfied to meet the POIP condition when the priority vector is derived using the geometric mean method.
Subsequently, Brunelli and Cavallo \cite{Brunelli2020b} suggested an optimization model for obtaining the closest PCM that satisfies  the index-exchangeability condition (see model (\ref{Bru2020}) in the Online Supplement).
However, even when the PCM meets the index-exchangeability condition, the priority vector obtained through existing prioritization methods (such as EM, LLSM, and ILSDM) may still violate the POIP condition.

Mazurek and Ram{\'\i}k \cite{Mazurek2019} introduced an optimization-based model to derive the priority vector considering the POIP condition  (refer to model (\ref{cp3ipops21}) in the Online Supplement).
However, this model presents two significant drawbacks:
first, the feasible solution set of model  (\ref{cp3ipops21}) is empty if  $\bf{A}$ violates the index-exchangeability condition;
second, the model is a non-convex optimization model, which may pose challenges in obtaining the optimal solution.

Even though reducing inconsistency does not necessarily get an answer closer to the `real' priority vector in some cases \cite{saaty1977,kazibudzki2019},
Kazibudzki \cite{kazibudzki2016} has proven through Monte Carlo simulations that high inconsistency may lead to erroneous choices of alternative solutions.
Additionally, the PCMs enable to comply with COP by improving the consistency among the pairwise comparisons \citep{kulakowski2019}.
As a result, employing methods aimed at reducing inconsistency and intransitivity \citep{WuZhibin2021Mtac,aguaron2021,brunelli2020,Xu2021,Liu2021,zhang2021consistency,li2022consensus,xu2023local} can be considered a viable alternative.

\subsection{Proposed work}
\label{subp}
Meeting the COP can not only reduce communication costs in the decision-making process but also confer intangible benefits, such as giving DMs a sense of fairness.
While there have been studies examining COP, no research has focused on methods to minimize or entirely avoid COP violations when the PCM fails to meet the index-exchangeability condition.
Therefore, we develop optimization models to determine the priority vector that satisfies COP requirements when faced with inconsistent PCMs.
The contribution of this paper is summarized as follows: 

\begin{itemize}
\setlength{\itemsep}{0pt}
\setlength{\parsep}{0pt}
\setlength{\parskip}{0pt}
\item[(1)]
Detection of COP Violations: As demonstrated by Bana e Costa and Vansnick \cite{BanaECostaCarlosA2008}, even when there exists a priority vector that meets the COP, EM and LLSM may fail to generate a priority vector from the acceptable PCM satisfying COP.
This indicates that the inherent flaws of EM and LLSM can lead to COP violations, and the eigenvalue-based consistency check fails to detect and warn of this phenomenon.
 Consequently, the method for detecting COP violations should be independent of the prioritization methods.
Building upon the work of Cavallo and D'Apuzzo \cite{cavallo2020}, we found that a priority vector exists that meets the POIP conditions when the PCM satisfies the index-exchangeability condition (see Proposition \ref{theoremiff}).
Therefore,  the index-exchangeability condition can be considered  the necessary and sufficient condition  for detecting the POIP violations   based on DM's pairwise comparison judgments.

\item[(2)]
Prioritization accounting for the COP:
In normative decision-making using pairwise comparison method, 
the priority vector's compliance with the COP depends on the reliability of the prioritization methods and the DM's rationality.
Although substantial research exists on prioritization methods \citep{Saaty1980,Crawford1985,CHOO2004,siraj2012,kulakowski2015,kazibudzki2022,kazibudzki2021,Zhang2021,Faramondi2020,yuen2024,wang2021,wang2021b},
there has been limited exploration of priority vectors considering POIP conditions.
To address this gap, we propose a two-stage programming model, MNVDM (see (\ref{cp3ipops2common})), to derive the priority vector.
The first stage reduces POIP violations, while the second stage minimizes deviations between the given and estimated PCMs.
MNVDM derives a priority vector that meets POIP conditions if the PCM satisfies the index-exchangeability condition. 
If not, it produces a vector with minimal POIP violations.
Compared to the method by Mazurek and Ram{\'\i}k \cite{Mazurek2019}, MNVDM offers two key advantages:
it  is a mixed-integer convex programming problem that can be quickly solved to obtain the optimal solution;
it serves as a general prioritization method, unlike Mazurek and Ram{\'\i}k's model \cite{Mazurek2019},  which requires the PCM to satisfy the index-exchangeability condition.

\item[(3)]
Handling unavoidable COP violations:
COP violations are unavoidable when the index-exchangeability condition is not met. 
In such instances, it is advisable to engage with the DMs to amend any unreasonable judgments and thereby obtain a reliable priority vector.
This paper introduces an optimization model designed to ensure COP by adjusting preferences with minimal information loss (refer to model (\ref{cp33poip12})).
Here, information loss is quantified by the number of adjusted preferences and their deviations.
Compared to  inconsistency reduction methods \citep{WuZhibin2021Mtac,aguaron2021,brunelli2020,Xu2021,Liu2021} and
the elimination method for the  index-exchangeability violation \cite{Brunelli2020b}, 
model (\ref{cp33poip12}) guarantees that both an acceptable level of inconsistency and the COP of the revised PCM are achieved.
\end{itemize}

\section{Preliminaries}\label{sec3}
This section presents some basic concepts: consistency indices, prioritization
methods and COP.

\begin{definition}\citep{Saaty1980}\label{consistencysaaty}
A PCM $ \mathbf{A}=(a_{ij})_{n\times n} $ is   consistent if
\begin{equation} \label{cardinal consistency}
a_{ij}=a_{ik} \cdot a_{kj}, \quad i,j\in {N}.
\end{equation}
\end{definition}

Actually, (\ref{cardinal consistency})  is the same as the following equation:
\begin{equation}
a_{i j}=w_{i} / w_{j}, \quad i,j\in {N}.
\end{equation}

However, because of cognitive limitations and a lack of information, the PCMs offered by DMs in real-world decision-making situations may be inconsistent or even intransitive. Thus,
Saaty \cite{Saaty1980} suggested the consistency ratio (CR):

\begin{equation} \label{CR}
\text{CR}(\mathbf{A})=\frac{\lambda_{max}(\mathbf{A})-n}{(n-1)\text{RI}},
\end{equation}
where $\lambda_{max}(\mathbf{A})$ is the maximum eigenvalue of $\mathbf{A}$, RI  is the random index presented in Table \ref{tableRI}.
According to  Saaty \cite{Saaty1980},   $\mathbf{A}$ is acceptable when $\text{CR}\leq 0.1$.

\begin{table}[h]
\begin{center}
\caption{The $RI$ provided by \cite{Aguarón2003}.}
\small
\newcommand{\tabincell}[2]{\begin{tabular}{@{}#1@{}}#2\end{tabular}}
\begin{tabular}{lllllllllll}%
\hline
$n$& 3& 4& 5& 6& 7& 8& 9\\
\hline
RI &0.525& 0.882& 1.115& 1.252 &1.341& 1.404& 1.452\\
\hline
\end{tabular}
\label{tableRI}
\end{center}
\end{table}

\begin{definition} \citep{Saaty1980} The priority vector $\mathbf{w}$ of $\mathbf{A}$ by the eigenvector method  (EM) is determined by
\begin{equation}
\left\{\begin{array}{l}
\mathbf{A w}=\lambda_{\max } \mathbf{w} \\
\mathbf{w}^{T} \mathbf{1}=1
\end{array}\right.
\end{equation}
where $\lambda_{\max }$ is the principal  eigenvalue of $\mathbf{A}$, and $\mathbf{1}=(1, \ldots, 1)^{T}$.
\end{definition}

\begin{definition}\citep{Crawford1985} The priority vector $\mathbf{w}$  of $\mathbf{A}$ by the Logarithmic Least Squares Method (LLSM) is determined by
\begin{equation}\label{LLSMC}
\left\{\begin{array}{l}
\min \sum\limits_{i= 1}^{n-1}\sum\limits_{j=i+1}^{n} (\ln (a_{ij} ) - \ln ({w_i}) + \ln ({w_j}))^2\\
\text { s.t. } \sum_{i=1}^{n} w_{i}=1, w_{i} > 0
\end{array}\right.
\end{equation}
\end{definition}

The LLSM is also called the Geometric Mean Method (GMM), because the optimal solution to the model (\ref{LLSMC}) can be expressed as follows:
\begin{equation}\label{wigci}
w_{i}=\frac{\sqrt[n]{\prod_{j=1}^{n} a_{i j}}}{\sum_{i=1}^{n} \sqrt[n]{\prod_{j=1}^{n} a_{i j}}},  i \in {N}.
\end{equation}

LLSM is gaining increasing popularity due to its sound axiomatic arguments \citep{csato2021,csato2019,bozoki2018}.
Aguar{\'{o}}n and Moreno-Jim{\'{e}}nez \cite{Aguarón2003}   introduced the inconsistency measure associated with LLSM, referred to as the Geometric Consistency Index (GCI).
They also provided the threshold  for the GCI, making this measure practical for application.

\begin{definition}\label{defGCI} \citep{Aguarón2003}
The Geometric Consistency Index (GCI) is defined as:
\begin{equation}\label{DongGCI}
\begin{split}
&\text{GCI}(\mathbf{A})\\
& = \frac{2}{{{(n-1)(n-2)}}}\sum\limits_{i= 1}^n\sum\limits_{j = i+1}^n (\log (a_{ij} ) - \log ({w_i}) + \log ({w_j}))^2,
\end{split}
\end{equation}
where ${w_i}$ is determined by Equation (\ref{wigci}).
\end{definition}

If $\text{GCI}(\mathbf{A}) = 0$, then $\mathbf{A}$ is  a consistent PCM.
A lower value of $\text{GCI}(\mathbf{A})$ indicates a lower inconsistency level of the PCM $\mathbf{A}$.
According to Aguar{\'{o}}n and Moreno-Jim{\'{e}}nez \cite{Aguarón2003}, the inconsistency level of $\mathbf{A}$ is acceptable when $\text{GCI}(\mathbf{A}) \leq\overline{\text{GCI}}$.
The value of $\overline{\text{GCI}}$ is as follows:
$\overline{\text{GCI}}=0.31$ when $n=3$; $\overline{\text{GCI}}=0.35$ when $n=4$ and $\overline{\text{GCI}}=0.37$ when $n>4$.

\begin{definition} \label{defLSDM}
\cite{kazibudzki2022,kazibudzki2021}
The priority vector $\mathbf{w}$  of $\mathbf{A}$ by  the Logarithmic Squared Deviations Method (LSDM) is  determined by

\begin{equation}\label{LSDM}
\left\{\begin{array}{l}
\min \sum\limits_{i=1}^{n} \ln ^{2}\left(\sum_{j=1}^{n}\left(a_{i j} w_{j} / (n w_{i})\right)\right)\\
\text { s.t. } \sum\limits_{i=1}^{n} w_{i}=1, w_{i} > 0
\end{array}\right.
\end{equation}
\end{definition}
The objective of model (\ref{LSDM}) can be considered as an inconsistency index, called the inconsistency index of logarithmic squared deviations method (ILSDM).
For the limited space, the other two recently proposed methods minimal error method (MEM) \cite{kulakowski2019} and ARDI \cite{Zhang2021} are presented in the Online Supplement.

\begin{definition}\citep{WuZhibin2021Mtac}
\label{de4}
Let $\mathbf{A} = {({a_{ij}})_{n \times n}}$ be a PCM, it holds for
  $i\in N, i<k, i<j$:

\begin{equation}\label{tuordinalconsistency}
\begin{array}{llllll}
\text{Condition (1)}: & a_{ik}\geq 1 & and & a_{kj} > 1 \Rightarrow a_{ij}> 1;\\
\text{Condition (2)}: & a_{ik} > 1& and & a_{kj}\geq1 \Rightarrow a_{ij}> 1;\\
\text{Condition (3)}: & a_{ik}=1 & and & a_{kj} = 1 \Rightarrow a_{ij}= 1.\\
 \end{array}
\end{equation}
then $\mathbf{A}$ is transitive (ordinally consistent).
\end{definition}

\begin{definition}Let $\mathbf{A} $ be a PCM satisfying (\ref{tuordinalconsistency}), its ranking vector $\textbf{r} = ({r_1},{r_2},\ldots,{r_n})^T$ is an actual ranking,
where
\begin{equation}
\begin{split}
r_i=&n-Card (a_{ij}|a_{ij}>1, j \in N, j\neq i)\\
    & -\frac{1}{2} Card(a_{ij}|a_{ij}=1,  j \in N, j\neq i),  \forall i \in N, \\
\end{split}
\end{equation}
where $Card(C)$ denotes the number of elements in the set  $C$.

\end{definition}

\begin{example}\label{eample1}
The above concepts are illustrated by the PCM proposed by Aguar{\'o}n et al. \cite{aguaron2021}.
\end{example}
\[\mathbf{A} =\left(\begin{array}{ccccccc}
1     & 4.35  & 6     & 8.05 \\
0.23  & 1     & 4     & 6 \\
1/6  &  1/4  & 1     & 3.48 \\
0.12  & 0.166667 & 0.29  & 1 \\
\end{array}\right)\]

$\text{GCI}(\mathbf{A})=0.3445<0.35$, and $\mathbf{A}$ satisfies the conditions in (\ref{tuordinalconsistency}). Thus,  $\mathbf{A}$ is acceptable and transitive.
Since $\mathbf{A}$ is transitive, its actual ranking is $\mathbf{r}=(1,2,3,4)^T$.

\begin{definition}\citep{cavallo2020}
Let $\mathbf{A} = {({a_{ij}})_{n \times n}}$ be a PCM.  If the following conditions are satisfied for all $\forall i, k \in N, i<j, k<l$:

\begin{equation}
\label{theorem2}
\begin{array}{llllll}
\text{Condition (1)}: & a_{ij}> a_{kl}  \Leftrightarrow a_{ik}> a_{jl}, \\
\text{Condition (2)}: & a_{ij}= a_{kl}  \Leftrightarrow a_{ik}= a_{jl}.\\
 \end{array}
\end{equation}
then $\mathbf{A}$ satisfies the index-exchangeability condition.
\end{definition}

\begin{definition}\label{rov}\citep{BanaECostaCarlosA2008}
Let $\mathbf{A}$ = ${({a_{ij}})_{n \times n}}$ and  $\textbf{w}$ = $({w_1},{w_2},\ldots,{w_n})^T$ be as before. It holds for $\forall i \in N, i<j$:
\begin{equation}
\label{depop}
\begin{array}{llllll}
\text{Condition (1)}: & a_{ij}> 1  \Leftrightarrow w_i>w_j, \\
\text{Condition (2)}: & a_{ij} = 1 \Leftrightarrow w_i=w_j.\\
 \end{array}
\end{equation}
Then, $\mathbf{w}$ satisfies the preservation of order preference (POP) conditions.
\end{definition}

\begin{definition}\citep{BanaECostaCarlosA2008}
\label{defPOIP}
Let $\mathbf{A} = {({a_{ij}})_{n \times n}}$ and  $\mathbf{w}= ({w_1},{w_2},\ldots,{w_n})^T$ be as before.
It holds for $\forall i, k \in N, i<j, k<l$:
\begin{equation}
\label{de4eq}
\begin{array}{llllll}
\text{Condition (1)}: & a_{ij}> a_{kl}  \Leftrightarrow w_i/w_j>w_k/w_l,\\
\text{Condition (2)}: & a_{ij}= a_{kl}  \Leftrightarrow w_i/ w_j=w_k/ w_l.\\
\end{array}
\end{equation}
Then, $\mathbf{w}$ satisfies the preservation of order of intensity preference   (POIP) condition.
\end{definition}

Specially, when $k=l$, Equation (\ref{de4eq}) is equivalent to Equation (\ref{depop}). 
Thus, $\mathbf{w}$ satisfying POIP  implies that $\mathbf{w}$ meets the POP conditions.
Conversely, if any of the conditions in Equations (\ref{depop}) or (\ref{de4eq}) are not satisfied, then $\mathbf{A}$ violates the COP. 
This paper refers to this phenomenon  as the `COP violation'.
The COP is exactly satisfied when $\mathbf{A}$ is consistent. But what happens if  $\mathbf{A}$ is inconsistent?

\section{Detecting the violations of COP}
\label{sec4}
This section analyses the relationship between transitivity, POP, and POIP when $\mathbf{A}$ is inconsistent.

\begin{proposition}\label{lemma1}\citep{cavallo2016}
There is a priority vector $\textbf{w}$ satisfying the POP conditions if and only if  $\mathbf{A}$ is transitive.
\end{proposition}

For a given PCM, it is crucial to determine whether there exists  a  priority vector $\mathbf{w}$ satisfying the POIP conditions.
Existing studies that aim to detect POIP violations predominantly rely on prioritization methods \citep{BanaECostaCarlosA2008}.
However, some of these prioritization methods can themselves lead to POIP violations.
Thus, controlling the influence of prioritization methods is essential when investigating the causes of POIP violations. 
Cavallo and D'Apuzzo \cite{cavallo2020} have proposed a necessary condition so that the priority vector preserves the POIP conditions.
Based on their work, we have identified a necessary and sufficient condition for the POIP conditions.

\begin{proposition}\label{theoremiff}
There is a $\mathbf{w}$ satisfying the POIP conditions if and only if  $\mathbf{A} $ satisfies the index-exchangeability condition.
\end{proposition}

Determining whether a PCM meets the index-exchangeability condition is solely dependent on the inconsistency level of the preferences and is independent of the prioritization method. 
Consequently, the index-exchangeability condition can be employed to detect violations as outlined in Proposition \ref{theoremiff}. 
The specific roles of the index-exchangeability condition are as follows:

\begin{itemize}
\setlength{\itemsep}{0pt}
\setlength{\parsep}{0pt}
\setlength{\parskip}{0pt}
\item[(1)]  The index-exchangeability condition can be  employed  to ascertain whether a priority vector $\textbf{w}$ that meets the COP exists, solely based on the pairwise comparisons in $\mathbf{A}$. If $\mathbf{A}$ violates the index-exchangeability condition, no method can yield a priority vector $\textbf{w}$ that satisfies the COP. 
    
\item[(2)] 
The index-exchangeability condition can be utilized to identify the causes of COP violations. 
If  $\mathbf{A}$ satisfies the index-exchangeability condition but the priority vector $\textbf{w}$ obtained through a particular prioritization method violates the COP, then the violations can be attributed to the prioritization method. 
Conversely, if $\mathbf{A}$  violates the index-exchangeability condition, as indicated by Proposition \ref{theoremiff}, it implies that no prioritization method can produce a priority vector $\textbf{w}$ that satisfies the COP,  and thus, the violations are due to the inconsistency in  $\mathbf{A}$.

\end{itemize}

While several studies have demonstrated that priority vectors are  more probable to satisfy the POIP conditions when PCMs exhibit a lower inconsistency level \citep{JiriMazurek2020,kulakowski2019},  it is noteworthy that PCMs meeting the index-exchangeability condition can still possess a high level of inconsistency.
Consider the following example.

\begin{example}\label{eampleindea} Let $\mathbf{A}$ be a PCM with five alternatives $\{x_1,x_2,x_3,x_4,x_5\}$.
\end{example}
\[\mathbf{A} =\left(\begin{array}{ccccccc}
1   & 6   & 7   & 8   & 9 \\
1/6 & 1   & 6   & 7   & 8 \\
1/7 & 1/6 & 1   & 6   & 7 \\
1/8 & 1/7 & 1/6 & 1   & 6 \\
1/9 & 1/8 & 1/7 & 1/6 & 1
\end{array}\right)\]

$\mathbf{A}$  satisfies the index-exchangeability condition, that is, there exists a priority vector satisfying the POIP conditions. However, the inconsistency level is pretty high, $\text{GCI}(\mathbf{A})=0.4587$.

\begin{figure}[h]
		\centering
		\includegraphics[scale=0.4]{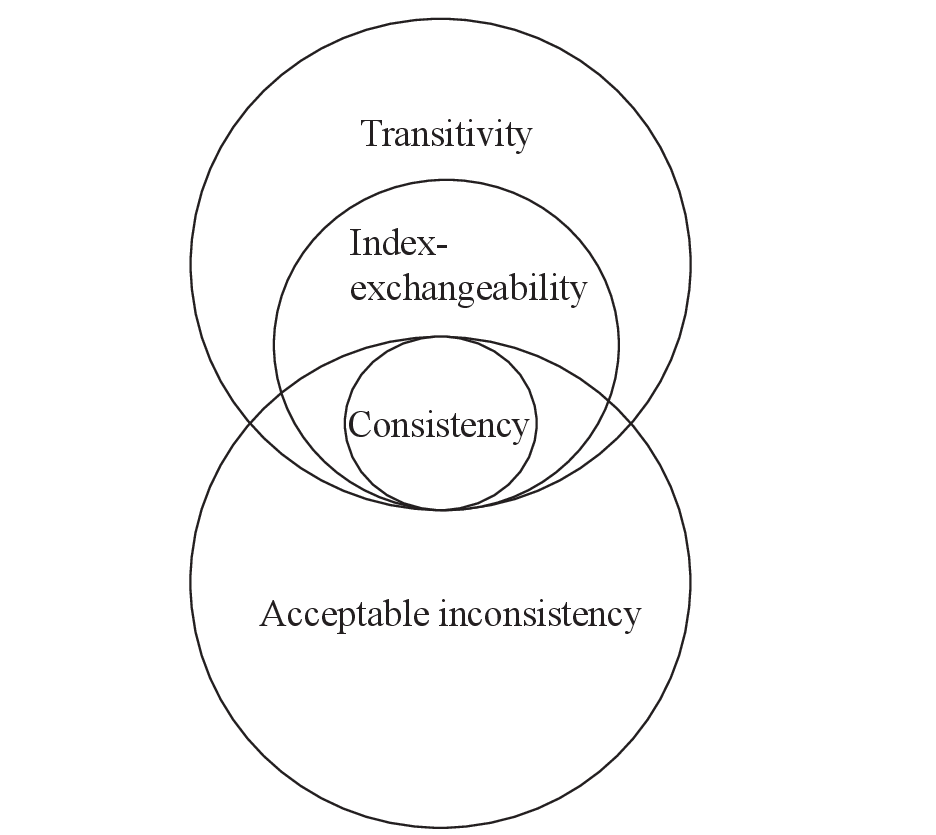}
		\caption{The relationship between consistency measures}
		\label{fig:1}
\end{figure}

By combining Proposition \ref{lemma1}  and Proposition \ref{theoremiff}, we can obtain the relationship between consistency, transitivity, and index-exchangeability as Fig. \ref{fig:1}.

\begin{example}\label{eample11} Let $\mathbf{A}$ be a PCM with four alternatives $\{x_1,x_2,x_3,x_4\}$.
\end{example}
\[\mathbf{A} =\left(\begin{array}{ccccccc}
1   & 2   & 4   & 9 \\
1/2 & 1   & 3   & 7 \\
1/4 & 1/3 & 1   & 5 \\
1/9 & 1/7 & 1/5 & 1
\end{array}\right)\]

$\text{GCI}(\mathbf{A})$=0.0658, $\text{CR}(\mathbf{A})=0.0377$,  $\mathbf{A}$ is transitive but inconsistent, $\mathbf{r}=(1,2,3,4)^T$.
To  determine whether there is a  $\mathbf{w}$ satisfying POIP conditions by  Proposition \ref{theoremiff}, it is necessary to examine the relationship between every pair of elements in the upper triangle of the PCM.
This entails considering a total of $C_{n(n-1)/2}^2$ cases, where $n$ represents the rank of the PCM. 
For this example where \( n = 4 \), all 15 cases (as detailed in the Online Supplement) conform to the index-exchangeability condition described in Equation (\ref{theorem2}). Consequently, there exists a priority vector that satisfies the POIP conditions.
For example,  $\textbf{w} = (0.6456,	0.2582,	0.0861,	0.0101)^T$ satisfies the POIP conditions.
The PCM $\textbf{W}=({\frac{w_i}{w_j}})_{4 \times 4}$ estimated by $\textbf{w}$ is as follows:
\[\mathbf{W} =\left(\begin{array}{ccccccc}
1 & 2.5004 & 7.4983 & 63.9208 \\
0.3999 & 1 & 2.9988 & 25.5644 \\
0.1334 & 0.3335 & 1 & 8.5248  \\
0.0156 & 0.0391 & 0.1173 & 1
\end{array}\right)\]

For Example \ref{eample11}, we employ five methods to derive the priority vector of $\mathbf{A}$, with the results presented in TABLE \ref{5me}. 
Using EM, LLSM, LSDM and ARDI, we obtain $w_1/w_3>w_3/w_4$. 
However, the corresponding pairwise comparison values are $a_{13} = 4$ and $a_{34} = 5$, which implies  $a_{13}=4<a_{34}=5$. 
Consequently, the priority vectors derived by EM, LLSM, LSDM, and ARDI do not satisfy the POIP conditions in this example.

Conversely, the priority vector obtained by MEM \cite{kulakowski2019} does meet the POIP conditions. 
This observation could be attributed to the capability of method  MEM \cite{kulakowski2019} to reduce the likelihood of POIP violations in nearly consistent PCMs. Therefore, based on this example, we can conclude that POIP violations may arise from two primary factors: the prioritization methods used and inconsistencies in the judgments.

\begin{table}[H]
\centering
\caption{The priority vectors of Example \ref{eample11} by different methods.}
\begin{tabular}{lccc}
\hline
Methods & $\textbf{w}$ & $w_1/w_3$        & $w_3/w_4$        \\\hline
EM \citep{Saaty1980}      & (0.5048, 0.3122,   0.1414,  0.0416)$^T$ & 3.5696 & 3.3985 \\
LLSM \citep{Crawford1985}    & (0.5063, 0.3129,   0.1396,  0.0413)$^T$ & 3.6257 & 3.3847 \\
LSDM \cite{kazibudzki2022}    & (0.5048, 0.3123,   0.1413,  0.0416)$^T$ & 3.5718 & 3.3983 \\
MEM \cite{kulakowski2019}      & (0.4849, 0.3276,   0.1476,  0.0399)$^T$ & 3.2863 & 3.7004 \\
ARDI \cite{Zhang2021}    & (0.5490, 0.2745,   0.1373,  0.0392)$^T$ & 4.0000 & 3.5000 \\\hline
\end{tabular}
\label{5me}
\end{table}

When $\mathbf{A}$ satisfies the index-exchangeability   condition, there is a $\textbf{w}$ meeting the POIP conditions.
However, this priority vector $\textbf{w}$ is not unique, leading to potential significant deviations  between $\mathbf{A}$ and $\textbf{W}$. 
Thus, a prioritization method considering both POIP conditions and deviations in order to generate a unique $\textbf{w}$ is essential.

\section{The prioritization method considering COP} \label{sec5}
In this section, the minimal number of violations and  deviations method (MNVDM)  is proposed to obtain the priority vectors of  inconsistent PCMs.
The superiority of this model is then demonstrated through comparisons and examples.

\subsection{Proposed optimization model to derive the priority vector}
A `violation' occurs when a $\mathbf{w}$ violates the POIP conditions.
To be consistent with the POP violations formulated by Golany and Kress \cite{golany1993}, we compute the number of violations (NV) using  the following equation:

\begin{equation}
\text{NV}=\sum\limits_{i= 1}^n\sum\limits_{j=i+1}^n\sum\limits_{k=1}^n\sum\limits_{l=k+1}^n v_{i jkl},
\end{equation}
where
\begin{equation}\label{NR1}
v_{i jkl}= \begin{cases}
1 & \text { if }\left(a_{ij}> a_{kl}\right) \text { but }\left(w_i/w_j<w_k/w_l\right) \\
0.5 & \text { if }\left(a_{ij}> a_{kl}\right) \text { but}\left(w_i/w_j=w_k/w_l\right) \\
0.5 & \text { if }\left(a_{ij}= a_{kl}\right) \text { but }\left(w_i/w_j \neq w_k/w_l\right)\\
0 & \text { otherwise }\end{cases}.
\end{equation}

An important property of prioritization methods is  that the priority vector should align with  the DM's preferences.
NV, in other words, should ideally be minimized.
Despite $\mathbf{A}$ meeting the index-exchangeability condition, as shown in Example \ref{eample11}, some methods (EM, LLSM, LSDM and ARDI)  may violate the POIP conditions.
This occurrence can be attributed to the fact that these methods do not inherently consider the POIP conditions.
Therefore, we propose a prioritization method based on minimal NV, constructed as follows:

\textbf{Stage 1} Minimize NV
\begin{equation}\label{NR}
\left\{\begin{array}{l}
\min \text{NV}=\sum\limits_{i= 1}^n\sum\limits_{j=i+1}^n\sum\limits_{k=1}^n\sum\limits_{l=k+1}^n v_{i jkl}\\
\text { s.t. } \sum_{i} w_{i}=1, w_{i}>0
\end{array}\right.
\end{equation}

To represent the relationship between  $a_{ij}$ and $a_{kl}$, we introduce $l_{ijkl}^a$ and $e_{ijkl}^a$. For a given PCM  $\mathbf{A}$, the values of $l_{ijkl}^a$ and $e_{ijkl}^a$ are known and computed as follows:
\begin{equation}\label{lea22}
\begin{array}{lllll}
l_{ijkl}^a=\left\{\begin{array}{ll}1, &\text{if} \quad  a_{ij}>a_{kl}\\0, &\text{otherwise}\\ \end{array} \right.,
& e_{ijkl}^a=\left\{\begin{array}{ll}1, &\text{if} \quad  a_{ij}=a_{kl}\\0, &\text{otherwise}\\ \end{array} \right.
\end{array}
\end{equation}

Binary variables   $l_{ijkl}^w$ and $e_{ijkl}^w$ are used to represent the value relationship between
$w_i/ w_j$ and $ w_k/w_l$, the relationship is as follows:
\begin{equation}\label{wwaa}
\begin{array}{lllll}
l_{ijkl}^w=\left\{\begin{array}{ll}1, &\text{if}  \quad \frac{w_i}{ w_j} > \frac{w_k}{w_l} \\0, &\text{otherwise}\\ \end{array} \right.,
&e_{ijkl}^w=\left\{\begin{array}{ll}1, &\text{if} \quad  \frac{w_i}{ w_j} = \frac{w_k}{w_l} \\0, &\text{otherwise}\\ \end{array} \right.
\end{array}
\end{equation}

By combining (\ref{NR1}), (\ref{lea22}) and (\ref{wwaa}), NV is equal to the following equation (refer to the Online Supplement  for the detailed calculation process):
\begin{equation} \label{xuci}
\begin{split}
\text{NV}=  & \sum\limits_{i= 1}^n\sum\limits_{j=i+1}^n\sum\limits_{k=1}^n\sum\limits_{l=k+1}^n
      (l_{ijkl}^a(1-l_{ijkl}^w-\frac{1}{2}e_{ijkl}^w)\\
    & + \frac{1}{2}e_{ijkl}^a(1-e_{ijkl}^w)).\\
\end{split}
\end{equation}

By Proposition \ref{theoremiff},  model (\ref{NR}) can get a priority vector satisfying POIP conditions when $\mathbf{A}$ satisfies the conditions in (\ref{theorem2}). For all $\forall i, k \in N, j>i, l>k$, model (\ref{NR}) can be mathematically expressed as the following model:

\begin{equation}
\label{cp3poip}
  \begin{array}{lll}
  \min \quad \text{NV}= \sum\limits_{i= 1}^{n-1}\sum\limits_{j=i+1}^{n}\sum\limits_{k=1}^{n-1}\sum\limits_{l=k+1}^n
      (l_{ijkl}^a(1-l_{ijkl}^w-\frac{1}{2}e_{ijkl}^w)\\
  \quad\quad\quad\quad\quad\quad   + \frac{1}{2}e_{ijkl}^a(1-e_{ijkl}^w))\\
  s.t.
    \left\{ \begin{array}{lll}
    -w_i/w_j+w_k/w_l \\- M(1 - l_{ijkl}^w)+\varepsilon \leq 0,    & (\ref{cp3poip}-1)\\
    w_i/w_j- w_k/w_l- Ml_{ijkl}^w \leq0, & (\ref{cp3poip}-2)\\
     -w_i/w_j+ w_k/w_l- M(1 - e_{ijkl}^w) \leq 0, & (\ref{cp3poip}-3)\\
     w_i/w_j- w_k/w_l- M(1 - e_{ijkl}^w) \leq 0,   & (\ref{cp3poip}-4)\\
     w_i/w_j- w_k/w_l- \\
     M(l_{ijkl}^w+e_{ijkl}^w) +\varepsilon \leq 0,  & (\ref{cp3poip}-5)\\
     l_{ijkl}^w+e_{ijkl}^w \leq 1, & (\ref{cp3poip}-6)\\
      \sum_{i} w_{i}=1, w_{i}>0,  & (\ref{cp3poip}-7)\\
      l_{ijkl}^w, e_{ijkl}^w \in \{0,1\},  & (\ref{cp3poip}-8)\\
  \end{array} \right.
  \end{array}
\end{equation}

For model (\ref{cp3poip}),  $l_{ijkl}^w$, $e_{ijkl}^w$, and $w_i$ are decision variables, $M$ is a big value and $\varepsilon$ is a positive number close to zero.
Constraints (\ref{cp3poip}-1) and (\ref{cp3poip}-2) ensure $l_{ijkl}^w=1 \Leftrightarrow  w_i/ w_j > w_k/w_l$;
constraints (\ref{cp3poip}-3) to (\ref{cp3poip}-5) express $e_{ijkl}^w=1 \Leftrightarrow  w_i/ w_j = w_k/w_l$;
for  a given $\mathbf{A}$, the values of $l_{ijkl}^a$ and $e_{ijkl}^a$ are known and determined by (\ref{lea22}).

\begin{proposition}\label{propmlp}
Model (\ref{cp3poip}) can be transformed into an mixed integer programming model as model (\ref{cp33poip}) (refer to the Online Supplement).
\end{proposition}

\textbf{Stage 2} Minimize deviations $\cal{D}(\mathbf{A}, \mathbf{W})$

Given that the optimal solution of model (\ref{cp33poip}) is not unique, the deviations between $\mathbf{A}$ and $\textbf{W}$ could be significant.
 Hence, this paper proposes a secondary objective aimed at minimizing these deviations after achieving the minimal NV. 
While AHP scholars generally acknowledge certain flaws in the EM, there is still no consensus on the most appropriate deviation function $\cal{D}$. Consequently, this paper investigates five different methods, two of which are classic approaches: EM \citep{Saaty1980}  and LLSM \citep{Crawford1985}. 
The other three methods have been proposed in the recent years: LSDM \cite{kazibudzki2022}, MEM \cite{kulakowski2019} and ARDI \cite{Zhang2021}. These methods share a common framework, termed the Minimal Number of Violations and Deviations Method (MNVDM):

\begin{equation}
\label{cp3ipops2common}
  \begin{array}{l}
  \min \quad {\cal{D}}(\mathbf{A}, \mathbf{W})\\
  s.t.
    \left\{ \begin{array}{ll}
    \text{NV}= \text{NV}^*\\
    \text{constraints (\ref{cp3poip}-1)-(\ref{cp3poip}-8)} \\
  \end{array} \right.
  \end{array}
\end{equation}
where $\text{NV}^*$ is the optimal value of model (\ref{cp33poip}).

Take LLSM as an example, the second stage  is to minimize the objective $\quad {\cal{D}}(\mathbf{A}, \mathbf{W}) = \frac{2}{{{(n-1)(n-2)}}}\sum\limits_{i= 1}^{n-1}\sum\limits_{j =i+1}^n (\log (a_{ij} ) - y_i + y_j)^2$. 
Then we denote this two stage model as the minimal number of violations based on the logarithmic least square method (MNVLLSM) model.

\emph{MNVLLSM}:
\begin{equation}
\label{cp3ipops2}
  \begin{array}{l}
  \min \quad {\cal{D}}(\mathbf{A}, \mathbf{W}) = \frac{2}{{{(n-1)(n-2)}}}\sum\limits_{i= 1}^{n-1}\sum\limits_{j =i+1}^n (\log (a_{ij} ) - y_i + y_j)^2\\
  s.t.
    \left\{ \begin{array}{ll}
    \text{NV}= \text{NV}^*\\
    \text{constraints (\ref{cp33poip}-1)-(\ref{cp33poip}-8)} \\
  \end{array} \right.
  \end{array}
\end{equation}
where $\text{NV}^*$ is the optimal value for model (\ref{cp33poip}).

Compared to  LLSM, the feasible region of $\textbf{w}$ in the MNVLLSM model is more constrained. 
When $\textbf{w}$ obtained by LLSM satisfies the COP, the MNVLLSM model degenerates into the LLSM and the deviation function $\cal{D}(\mathbf{A}, \mathbf{W})$ becomes equivalent to $\text{GCI}(\mathbf{A})$. 
Hence, ${\cal{D}}(\mathbf{A}, \mathbf{W})$ can be interpreted as a consistency measure incorporating the COP.

Similar to the unique solution provided by LLSM, the MNVLLSM also ensures a unique optimal solution. 
The objective function ${\cal{D}}(\mathbf{A}, \mathbf{W})$) in the MNVLLSM model is a quadratic convex function, and the constraints are all mixed-integer functions. 
Various exact algorithms, such as the branch and bound algorithm, are capable of obtaining the optimal solution for this type of model.
The total number of decision variables is $2 \times ({\frac{n(n-1)}{2}})^2+n = {\frac{(n(n-1))^2}{2}} +n$, where $n$ denotes the order of $\mathbf{A}$. 
The complexity of model (\ref{cp3ipops2}) escalates rapidly as $n$ increases. 
Nevertheless, for values of $3\leq n\leq 9$, the optimal solution of model (\ref{cp3ipops2}) can be obtained in a short time using existing optimization solvers, like Gurobi. 
The range of $n$ from 3 to 9 is primarily owing to the cognitive limitation that DMs cannot simultaneously compare more than seven objects (plus or minus two) \citep{miller1956}.

The two-stage models based on four other prioritization methods, specifically EM, MEM, LSDM and RADI are presented on the Online Supplement.
It is important to note that all five methods (MNVLLSM, MNVLSDM, MNVARDI, MNVMEM, and MNVEM) exhibit identical performance in minimizing the NV, as their first-stage models share the same objective to minimize violations.
Additionally, substituting the LLSM in model (\ref{cp3ipops2}) with other non-convex optimization-based prioritization methods like the least square method \cite{CHOO2004} or the least absolute error method \cite{CHOO2004}, transforms model (\ref{cp3ipops2}) into a mixed-integer non-convex optimization problem. These types of problems are notoriously difficult to solve, and finding optimal solutions can pose significant challenges.

\begin{remark}
In this paper, we incorporated constraints pertaining to the POIP conditions into the five prioritization methods: EM, LLSM, LSDM, MEM  and ARDI.
This  ensures that the priority vectors derived from the PCM exhibit the minimum number of POIP violations. 
The resultant models (MNVLLSM, MNVLSDM, MNVARDI, MNVMEM, MNVEM) are optimization models which can be efficiently solved using existing solvers.
Specifically,  MNVMEM is a mixed-integer programming  (MIP)  model, MNVLLSM is a mixed-integer quadratic programming (MIQP) model, and the other models (MNVLSDM,  MNVARDI, MNVEM) are mixed-integer nonlinear programming (MINP) problems and their relaxation forms are convex programming models. 
To further enhance solving efficiency, we utilized \texttt{Gurobi (version 10.1)}\footnote{\href{https://www.gurobi.com/products/gurobi-optimizer}{https://www.gurobi.com/products/gurobi-optimizer}} to solve both MNVLLSM and MNVMEM models. The nonlinear optimization solver, \texttt{Ipopt}\footnote{\href{https://github.com/coin-or/Ipopt}{https://github.com/coin-or/Ipopt}}, was employed to solve the other MINP models (MNVLSDM, MNVARDI, and MNVEM). All algorithms were implemented in \texttt{Matlab 2018b}, using \texttt{Yalmip}\footnote{\href{https://yalmip.github.io}{https://yalmip.github.io}} as the mathematical modeling software.

\end{remark}

\subsection{Numerical examples and comparisons}
\label{da}
In this section, we utilize Monte Carlo simulations to generate nearly consistent PCMs. This approach involves constructing a consistent PCM and introducing perturbations to its elements. It has been extensively employed in previous studies on PCMs \citep{kazibudzki2022,JiriMazurek2020,szadoczki2023,csato2024}.

\subsubsection{Dataset}
1000  PCMs  are  randomly generated for each situation where the values of $n$ is  3 to 9 using the Monte Carlo simulation method.
Take $n=5$ as an example to illustrate the process of generating random PCMs:

\begin{itemize}
\setlength{\itemsep}{0pt}
\setlength{\parsep}{0pt}
\setlength{\parskip}{0pt}
\item[(1)] Choose $n=5$ uniformly distributed random number $w_i$  from the interval [1,9].
\item[(2)] Compute the consistent PCM $\mathbf{A}=(a_{ij})$, where $a_{ij}=w_i/w_j$.
\item[(3)] For all $i\neq j$, if $a_{ij}\geq 1$, then  $a_{ij}$ is changed to  $\widetilde{a}_{ij}=a_{ij} \times \delta_{i j}$, where $\delta_{i j}$ is randomly chose from the interval [1,1.2].
    Then according to the reciprocality, $\widetilde{a}_{ji} =\widetilde{a}_{ij}$.
\item[(4)] Round  all  the upper triangle entries $\widetilde{a}_{ij}$ ($i<j$) to the Saaty's scale. 
The lower triangle entries $\widetilde{a}_{ij}$ ($i>j$) are computed by the reciprocity, $\widetilde{a}_{ij}\cdot \widetilde{a}_{ji}=1$.
\end{itemize}

\subsubsection{Comparative analysis}
\label{subscom}
To investigate the performance of MNVDM (model (\ref{cp3ipops2common})),  we compare it with the two most widely used methods (EM and LLSM), the recently proposed methods including  the model MEM \cite{kulakowski2019}, the ARDI model \cite{Zhang2021}, the LSDM \cite{kazibudzki2022}.
Because the NV will not be affected by the second stage objective function ${\cal{D}}(\mathbf{A}, \mathbf{W}) $ of MNVDM, that is, 
the NV obtained by these five methods (MNVLLSM, MNVLSDM, MNVARDI, MNVMEM, MNVEM) will be the same.
For this reason, we utilize MNVMEM in this section, as it is a MIP  model, which is more easily solvable compared to other MINP models (MNVEM, MNVLLSM, MNVLSDM, and MNVARDI).

The NV of PCMs in the Dataset from Section \ref{da} is calculated using all these methods, 
and the average number of violations are summarized in Table \ref{tab:simu}. Based on these results, we can draw the following conclusions:

\begin{itemize}
\setlength{\itemsep}{0pt}
\setlength{\parsep}{0pt}
\setlength{\parskip}{0pt}
\item[(1)] The proposed MNVDM (model (\ref{cp3ipops2common})) can generate significantly smaller NV compared to the other methods.
This is because the first stage of MNVDM minimizes the NV, whereas the other models only aim to minimize deviations.

\item[(2)] Regarding the performance of the existing methods (EM, LLSM, LSDM, MEM and ARDI) in avoiding POIP violations, 
EM is always dominated by LLSM and MEM \cite{kulakowski2019} under different values of $n$, which suggests that EM should be replaced by newly proposed methods.

\item[(3)]  As $n$ increases, the performance of LSDM on NV deteriorates. Specifically, LSDM performs similarly to LLSM and MEM \cite{kulakowski2019} when $n=3,4$; however, LSDM performs the worst when $n\geq 5$.
Conversely, shows improved performance on NV as  $n$ increases.
Nevertheless, within the MNVDM framework, the definition of ${\cal{D}}(\mathbf{A}, \mathbf{W})$ has no impact on NV.
For example, by replacing  LSDM with MNVLSDM for $n=9$,  NV will be reduced by more than 50\% (from 97.78 to 46.50).

\end{itemize}

\begin{table}[h]
    \centering
    \caption{The average number of violations under different $n$.}
\begin{tabular}{llllllll}
\hline
n      & 3    & 4    & 5    & 6     & 7     & 8     & 9     \\\hline
EM \cite{Saaty1980}    & 0.14 & 1.35 & 4.79 & 12.38 & 26.51 & 40.16 & 73.91 \\
LLSM \citep{Crawford1985}    & 0.07 & 1.11 & 4.06 & 11.59 & 24.89 & 38.84 & 72.05 \\
LSDM \cite{kazibudzki2022}    & 0.09 & 1.32 & 5.10 & 14.28 & 31.45 & 49.59 & 97.78 \\
MEM \cite{kulakowski2019}      & 0.10 & 1.27 & 4.25 & 11.80 & 25.26 & 39.96 & 72.28 \\
ARDI \cite{Zhang2021}    & 0.39 & 1.77 & 4.85 & 12.09 & 24.20 & 37.22 & 58.75 \\
MNVDM & \bf{0.04} & \bf{0.78} & \bf{2.68} & \bf{7.48}  & \bf{15.76} & \bf{25.90} & \bf{46.50} \\\hline
\end{tabular}
\label{tab:simu}
\end{table}

However, the improvement in order preservation performance by MNVDM is not free, as  it simultaneously increases the model's complexity and necessitates additional computational time.
For instance, while LLSM can arrive at a solution within 0.001 seconds, MNVLLSM requires more than 10 seconds for $n=9$.
The solution time  of MNVLLSM model various values of $n$  is shown in Table \ref{tabst}.
For all cases where $3\leq n\leq 9$, the MNVLLSM model can get the optimal solution within 30 seconds.
This demonstrates that the model is suitable for addressing real-world problems.

\begin{table}[H]
    \centering
    \caption{The solution time  of MNVLLSM model under different $n$ (seconds).}
    \begin{tabular}{lllllllll}
    \hline
        $n$     & 3    & 4    & 5    & 6    & 7     & 8    & 9 \\ \hline
        Max     & 0.09 & 0.11 & 0.37 & 0.52 & 5.35 & 9.46 & 31.47 \\
        Average & 0.04 & 0.05 & 0.08 & 0.17 & 0.48 & 1.30 & 11.37 \\
        Min     & 0.02 & 0.03 & 0.05 & 0.06 & 0.13 & 0.21 & 0.43 \\ \hline
    \end{tabular}
    \label{tabst}
\end{table}

\section{Procedure to eliminate violations}\label{sec6}
In this section, the POIP conditions are expressed as mixed-integer linear constraints. Then an optimization model is built to avoid POIP violations by minimizing the  number of adjusted preferences  and the amount of preference modification.  Finally,  numerical examples and comparisons are presented to demonstrate the model's feasibility and validity.

\subsection{Proposed optimization model to eliminate violations}

POIP violations are inevitable when the PCM fails to meet the index-exchangeability condition.
Due to the DM's limited global perception, a moderator is needed to provide suggestions in order to acquire more reliable results.
It is crucial that these suggestions closely align with the DM's initial preferences, meaning they should incur minimal information loss compared to the original preferences, as recommended by several AHP scholars \citep{saaty2003,Bozoki2015,WuZhibin2021Mtac,aguaron2021,li2019overview}.
Existing models generally have two main objectives for ensuring the acceptability of the revised PCM: minimizing deviations \citep{saaty2003,brunelli2020,Brunelli2020b} and minimizing the number of revised preferences \citep{Bozoki2015,WuZhibin2021Mtac}.
Thus, this paper considers both the  deviations and number of revised elements in finding the PCMs meeting the POIP conditions and the acceptable inconsistency level.
Example \ref{eampleindea} illustrates that a PCM meeting the index-exchangeability condition may still exhibit a high level of inconsistency.
Consequently, both inconsistency levels and POIP conditions must be considered in the model to effectively eliminate such violations.

Let $\mathbf{A}=(a_{ij})_{n\times n} $ and $\mathbf{\overline{A}}=(\overline{a_{ij}})_{n\times n}$ be the original PCM and revised PCM, respectively. Let
$\text{DS}_{[1/9,9]}=\{1/9, 1/8,1/7,1/6,1/5,1/4,1/3,1/2, 1,2,3,4,5,6,7,8,9\}$ be Saatty's  scale.
The  number of revised preferences (NRP) is computed by:
\begin{equation}
\text{NRP}=\sum\limits_{i = 1}^n\sum\limits_{j=i+1}^n \delta_{ij},
\end{equation}
where
\begin{equation*}
\delta_{ij}=\left\{\begin{array}{ll}
{1}, &\text{if  $\overline{a_{ij}}\neq {a_{ij}}$ }\\
0, &\text{otherwise.}\\
\end{array} \right.
\end{equation*}

The amount of changes (AOC) is computed by

\begin{equation}\label{secondmv}
\text{AOC}= \sum\limits_{i = 1}^{n}\sum\limits_{j=i+1}^n |\log a_{ij}-\log \overline{a_{ij}}|.
\end{equation}

The objective function is to minimize the information loss  measured by NPR and AOC, and is designed as follows
\begin{equation}
J=\alpha \text{NPR}+\text{AOC}
\end{equation}
where $\alpha\geq 0$, is used to measure the importance of NPR relative to AOC.
In this paper, we set $\alpha=1,000$.

Let $\bf{\overline{w}}$=$(\overline{w_1},\overline{w_2},\ldots,\overline{w_n})^T$ be the priority vector of $\mathbf{\overline{A}}$,
$\overline{y_i}=\log \overline{w_i}$, and $x_{ij}=\log \overline{{a}_{ij}}$.
Thus, for $\forall i, j,k,l \in N$, we can design the following common framework to eliminate POIP violations and reduce the inconsistency level:

\begin{equation}
\label{cp3poip11}
  \begin{array}{ll}
  \min \quad J=\alpha \text{NPR}+\text{AOC}\\
  s.t.\left\{ \begin{array}{lll}
    \overline{a_{ij}}> \overline{{a}_{kl}}  \Leftrightarrow \overline{w_i}/\overline{w_j} >\overline{w_k}/\overline{w_l} & (\ref{cp3poip11}-1)\\
    \overline{a_{ij}}= \overline{{a}_{kl}}  \Leftrightarrow \overline{w_i}/\overline{w_j} =\overline{w_k}/\overline{w_l} & (\ref{cp3poip11}-2)\\
    \sum\limits_{i = 1}^n \overline{w_i}=1 &  (\ref{cp3poip11}-3)\\
    |a_{ij}-\overline{a_{ij}}| \leq M \delta_{ij}  & (\ref{cp3poip11}-4)\\
     \cal{D}(\mathbf{\overline{A}},\overline{\mathbf{W}})\leq \overline{\text{CI}}  & (\ref{cp3poip11}-5)\\
     \overline{{a}_{ij}}{\times} \overline{{a}_{ji}}=1 & (\ref{cp3poip11}-6)\\
    \overline{{a}_{ij}}\in DS_{[1/9,9]}  & (\ref{cp3poip11}-7)\\
    \end{array} \right.
  \end{array}
\end{equation}

Constraints (\ref{cp3poip11}-1) and (\ref{cp3poip11}-2) the existence of a priority vector that meets the POIP condition, while constraint (\ref{cp3poip11}-3) normalizes this priority vector;
constraint (\ref{cp3poip11}-4)determines whether the preference $a_{ij}$ has been revised;
constraint (\ref{cp3poip11}-5) ensures that $\cal{D}(\mathbf{\overline{A}},\mathbf{\overline{W}})$ is below the threshold consistency index $\overline{\text{CI}}$;
the reciprocity is modelled as constraint (\ref{cp3poip11}-6);
constraint (\ref{cp3poip11}-7) requires that the revised preference $\overline{{a}_{ij}}$ belongs to the Saaty's scale.

The constraints (\ref{cp3poip11}-1), (\ref{cp3poip11}-2) and (\ref{cp3poip11}-6) are not directly tractable by the optimization solvers.
Therefore, we transform model (\ref{cp3poip11}) into a tractable form, designated as model (\ref{cp33poip12}),  detailed in the Online Supplement.

Model (\ref{cp3poip11}) considers both COP (or the index-exchangeability condition) and the consistency index  in the process of  inconsistency reduction, which provides more rational and coherent modification suggestions compared to the existing methods \citep{Brunelli2020b,WuZhibin2021Mtac,aguaron2021,saaty2003}. However, this model has the following two main limitations:

\begin{itemize}
\setlength{\itemsep}{0pt}
\setlength{\parsep}{0pt}
\setlength{\parskip}{0pt}
\item[(1)]
Transitivity is often regarded as fundamental to both normative and descriptive decision theories, underpinning the logical coherence of preferences \citep{regenwetter2011}. 
However, we acknowledge that in real-world decision-making, preferences frequently exhibit intransitivity. Factors such as context dependency, emotional influences, and bounded rationality contribute to these transitivity violations  \citep{tversky1969}.
Moreover, the concept of index-exchangeability is even stricter than transitivity, as depicted in Fig. \ref{fig:1}. 
Consequently, in some real-world decision-making scenarios, the preferences provided by the DM are more prone to infringing upon the index-exchangeability condition than the transitivity condition.
If the DM refuses to revise preferences that violate the index-exchangeability condition, the priority vector obtained by MNVDM will also contain many violations like the existing methods, and  model (\ref{cp3poip11})  will be infeasible.
Given these concerns, further empirical research is needed to determine the probability of such problems occurring and to better understand their implications. 

\item[(2)]
Analogous to the MNVDM model (\ref{cp3ipops2common}), the deviation function $\cal{D}(\mathbf{\overline{A}},\overline{\mathbf{W}})$ in constraint (\ref{cp3poip11}-5) can be substituted with  consistency indices that are based on the deviation minimization such as the GCI (see Definition \ref{defGCI}) and ILSDM (see Definition \ref{defLSDM}). 
Except constraint (\ref{cp33poip12}-8), all the other constraints in model (\ref{cp33poip12}) are linear.
 Consequently, the solvability of model (\ref{cp33poip12}) hinges entirely on the definition of the consistency index.
Although the inconsistency index is not our focus in this paper, we strongly recommend that  the consistency indices based on the deviation function $\cal{D}(\mathbf{\overline{A}},\overline{\mathbf{W}})$ be formulated as convex functions to facilitate computational efficiency.
\end{itemize}

\subsection{Numerical examples and comparisons}
\label{secp}

\subsubsection{A numerical example}
\begin{example}\label{ex7}
Let $\mathbf{A}$ be a PCM with eight alternatives $\{x_1,x_2,x_3,x_4, x_5,x_6,x_7,x_8\}$ which has been discussed by  Saaty \cite{saaty2003}.
\end{example}

\[\mathbf{A} =\left(\begin{array}{ccccccccccc}
 1     & 5     & 3     & 7     & 6     & 6     &  1/3  &  1/4 \\
     1/5  & 1     &  1/3  & 5     & 3     & 3     &  1/5  &  1/7 \\
     1/3  & 3     & 1     & 6     & 3     & 4     & 6     &  1/5 \\
     1/7  &  1/5  &  1/6  & 1     &  1/3  &  1/4  &  1/7  &  1/8 \\
     1/6  &  1/3  &  1/3  & 3     & 1     &  1/2  &  1/5  &  1/6 \\
     1/6  &  1/3  &  1/4  & 4     & 2     & 1     &  1/5  &  1/6 \\
    3     & 5     &  1/6  & 7     & 5     & 5     & 1     &  1/2 \\
    4     & 7     & 5     & 8     & 6     & 6     & 2     & 1     \\
  \end{array}\right)\]

$\text{GCI}(\mathbf{A})=0.5292>0.37$ indicates that $\mathbf{A}$ has a high level of inconsistency.
In $\mathbf{A}$, $a_{13}=3, a_{37}=6$, and $a_{71}=3$ imply that $\mathbf{A}$ is intransitive and contains a preference cycle ($x_{1} \succ x_{3} \succ x_{7} \succ x_{1}$).
Hence,  it is impossible to derive a priority vector that meets the POIP condition due to the DM's self-contradictory  preferences.
Therefore, suggesting  DMs  reconsider their  preferences is a better way.

In this example, model (\ref{cp33poip12}), subject to the constraint \( \text{GCI}(A) \le 0.37 \), is applied to generate modification suggestions. The revised PCM $\mathbf{\overline{A}}$ is presented as follows:
 \[\mathbf{\overline{A}} =\left(\begin{array}{ccccccccccc}
 1   & 5   & 3   & 7          & 6            & \textbf{4}   & 1/3          & 1/4          \\
1/5 & 1   & 1/3 & \textbf{4} & 3            & \textbf{1/2} & \textbf{1/6} & 1/7          \\
1/3 & 3   & 1   & 6          & \textbf{5}   & \textbf{3}   & \textbf{1/4} & 1/5          \\
1/7 & 1/4 & 1/6 & 1          & \textbf{1/2} & 1/4          & \textbf{1/8} & \textbf{1/9} \\
1/6 & 1/3 & 1/5 & 2          & 1            & \textbf{1/3} & \textbf{1/7} & \textbf{1/8} \\
1/4 & 2   & 1/3 & 4          & 3            & 1            & 1/5          & 1/6          \\
3   & 6   & 4   & 8          & 7            & 5            & 1            & 1/2          \\
4   & 7   & 5   & 9          & 8            & 6            & 2            & 1
  \end{array}\right)\]

$\mathbf{\overline{A}}$ satisfies the POIP conditions in (\ref{de4eq}),
 13 elements in the upper triangle are modified and marked in bold.
 Additionally, one judgment has been reversed, with $a_{37}=6$ now being $\overline{a_{37}}=1/4$.
Due to the presence of a preference cycle within $\mathbf{{A}}$, the judgment reversal   is inevitable in this example to break the cycle.
Similarly, Saaty \cite{saaty2003}  altered the preference direction of $a_{37}$ from 6 to 1/2 to mitigate the inconsistency of $\mathbf{{A}}$.
The directions of the other modified preferences remain consistent with the original preferences.

The actual ranking vector of  $\mathbf{\overline{A}}$ is $\mathbf{r}=(3,	6,	4,	8,	7,	5,	2,	1)^T$.
Table \ref{tab:example7r} shows the priority vectors derived using different methods.
All methods yield ranking vectors for $\mathbf{\overline{A}}$  that are identical to the actual ranking vector.
However, the priority vectors by existing methods (EM, LLSM, MEM, LSDM, and ARDI) exhibit  violations.
This discrepancy arises because these methods do not account for the COP in the prioritization process.
However, when we apply the MNVDM with different deviation functions (MNVEM, MNVLLSM, MNVMEM, MNVLSDM, and MNVARDI), all the resulting priority vectors  meet  the COP.
This outcome is due to the ability of the MNVDM to obtain a priority vector that satisfies the COP when the PCM satisfies the index-exchangeability condition.

It is noted that the  improved  performance of MNVDM  in COP  does not come without trade-offs. 
For instance, when replacing LSDM with MNVLSDM,  the deviation measure $\cal{D}(\mathbf{\overline{A}},\overline{\mathbf{W}})$ increases from  0.0415 to 0.1556. That means that   the MNVDM model compromises some cardinal deviations   to minimize violations.

\begin{table*}[!htb]
  \centering
  \small
  \caption{The priority vectors of $\mathbf{\overline{A}}$  using different methods for Example \ref{ex7}}
  \newcommand{\tabincell}[2]{\begin{tabular}{@{}#1@{}}#2\end{tabular}}
  \setlength{\tabcolsep}{1mm}{
    \begin{tabular}{lllcc}
     \hline
    \tabincell{l}{Prioritization Method} &  \tabincell{l}{$\cal{D}(\mathbf{\overline{A}}, \mathbf{W})$}  & {NV} & {\textbf{w}} \\
     \hline
    EM    & 0.0608 & 11    &$(0.1576,	0.0446,	0.0964,	0.0193,	0.0260,	0.0558,	0.2521,	0.3482)^T $\\
    LLSM  & 0.2221 &9   &$(0.1563,	0.0448,	0.0967,	0.0191,	0.0262,	0.0571,	0.2537,	0.3461)^T $\  \\
    MEM   &1.7818 &9.5   &$(0.1470,	0.0462,	0.0873,	0.0206,	0.0235,	0.0519,	0.2934,	0.3301)^T $\  \\
    LSDM   &0.0415 &9.5   &$(0.1574,	0.0446,	0.0964,	0.0193,	0.0260,	0.0560,	0.2524,	0.3478)^T $\  \\
    ARDI  &0.6140           &8     &$(0.1132,	0.0377,	0.0906,	0.0151,	0.0189,	0.0453,	0.2264,	0.4528)^T $\  \\
    MNVEM  & 0.1480   & 0   & $(0.1205,	0.0364,	0.0808,	0.0164,	0.0244,	0.0542,	0.2679,	0.3995)^T $ \\
    MNVLLSM &  0.2566 & 0   &$(0.1434,	0.0508,	0.1405,	0.0182,	0.0508,	0.0508,	0.1448,	0.4006)^T $\  \\
    MNVE    & 2.0801    & 0   & $(0.1253,	0.0418,	0.0869,	0.0201,	0.0290,	0.0603,	0.2607,	0.3760)^T $ \\
    MNVLSDM & 0.1556    & 0   & $(0.1220,	0.0380,	0.0827,	0.0174,	0.0257,	0.0560,	0.2658,	0.3923)^T $\\
    MNVARDI& 0.6813 & 0   &  $(0.1204,	0.0363,	0.0807,	0.0163,	0.0243,	0.0541,	0.2680,	0.3998)^T $ \\\hline
    \end{tabular}}%
  \label{tab:example7r}%
\end{table*}

Furthermore, it should be emphasized that the revised preferences given by  model (\ref{cp33poip12}) are not mandatory. If the DM doesn't agree with the proposed revisions, it is essential to engage in further dialogue and adjust other preferences based on the DM's feedback.
For example, if the DM firmly believes  that $a_{37}$ should be equal to 6, then we should  set $x_{37}=\log 6$ before applying model (\ref{cp33poip12}).

\subsubsection{Comparison with Brunelli and Cavallo \cite{Brunelli2020b}'s method}

Comparing to the model proposed by Brunelli and Cavallo \cite{Brunelli2020b} (see (\ref{Bru2020})), our model  (\ref{cp3poip11}) has three advantages:
the objective of model  (\ref{cp3poip11}) takes into account the NPR, less NPR will take less energy to communicate with DM;
since $\cal{D}(\mathbf{\overline{A}},\overline{\mathbf{W}})\leq \overline{\text{CI}}$, the revised PCM $\mathbf{\overline{A}}$ obtained by our model is below the inconsistency threshold;
the revised preference $\overline{a_{ij}}$ belongs to the Saaty's scale, which is very important in the feedback process  due to the specific semantic meaning associated with preferences on Saaty's scale. For example, $\overline{a_{16}}=3$ means that $x_1$ is slightly important than $x_6$.
\begin{example}\label{ex6}
Let $\mathbf{A}$ be a PCM with four alternatives $\{x_1,x_2,x_3,x_4\}$.

\end{example}
\[\mathbf{A} =\left(\begin{array}{ccccccc}
1   & 6   & 7   & 9 \\
1/6 & 1   & 6   & 8 \\
1/7 & 1/6 & 1   & 5 \\
1/9 & 1/8 & 1/5 & 1
  \end{array}\right)\]

$\mathbf{A}$ is transitive.
$\text{GCI}(\mathbf{A})=0.3781$  indicates that $\mathbf{A}$ is pretty inconsistent.
$a_{13}=7<a_{24}=8$ and $a_{12}=6>a_{34}=5$ mean that $\mathbf{A} $ violates the index-exchangeability condition.
To get the priority vector satisfying the POIP condition, we apply  the model proposed by Brunelli and Cavallo \cite{Brunelli2020b} (see (\ref{Bru2020})) and our model  (\ref{cp3poip11}) where the constraint (\ref{cp3poip11}-6) is replaced with $\text{GCI}(\mathbf{\overline{A}}) \leq 0.35$  to revise $\mathbf{A}$.
The results are as follows:
\[\mathbf{\overline{A}}_{\text{Brunelli}} =\left(\begin{array}{ccccccc}
1      & 6      & \textbf{7.3284} & 9      \\
1/6    & 1      & 6      & \textbf{6.6310} \\
0.1365 & 1/6    & 1      & 5      \\
1/9    & 0.1508 & 1/5    & 1
  \end{array}\right)\]
\[\mathbf{\overline{A}}_{\text{GCI}} =\left(\begin{array}{ccccccc}
1   & \textbf{4}   & 7   & 9 \\
1/4 & 1   & \textbf{3}   & 8 \\
1/7 & 1/3 & 1   & 5 \\
1/9 & 1/8 & 1/5 & 1
\end{array}\right)\]

Both the revised PCMs by  model (\ref{cp3poip11}) and model (\ref{Bru2020}) satisfy the index-exchangeability condition.
However, the revised PCM by  model (\ref{Bru2020}) has a  higher level of inconsistency level ($\text{GCI}(\mathbf{\overline{A}}_{\text{Brunelli}})=0.3747$ $>$ $\text{GCI}(\mathbf{\overline{A}}_{\text{GCI}})= 0.3449$) and the revised preferences are not the Saaty's scale.
For the revised PCM by  model (\ref{cp3poip11}), the revised PCM is acceptable inconsistent ($\text{GCI}(\mathbf{\overline{A}}_{\text{GCI}})= 0.3449<0.35$) and satisfies the index-exchangeability condition.

\subsubsection{Comparison with other inconsistency reducing methods}

To compare this method with existing inconsistency reduction techniques \citep{WuZhibin2021Mtac,aguaron2021,saaty2003},
1000 PCMs ($0.1< \text{CR}\leq 0.15$) are generated randomly for each situation where $n$  varies from 3 to 9  using the Monte Carlo simulation method mentioned in Section \ref{da}.
Initially, we applied the inconsistency reduction methods \citep{WuZhibin2021Mtac,aguaron2021,saaty2003} and model (\ref{cp3poip11}) where the constraint (\ref{cp3poip11}-6) is replaced with $\text{GCI}(\mathbf{\overline{A}}) \leq \overline{\text{GCI}}$   to reduce the inconsistency respectively. 
Subsequently,  the MNVLLSM model is utilized to derive the priority vector from the revised PCMs.

The comparative results of these methods are presented in TABLE \ref{sl1000}.
For the three inconsistency reduction methods \citep{WuZhibin2021Mtac,aguaron2021,saaty2003}, the values of NPR, AOC, and NV all increase as $n$ increases.
Specifically, the performance of the methods \citep{aguaron2021,saaty2003} in terms of NV, AOC, and NPR is quite similar.
This similarity arises because both methods  \citep{aguaron2021,saaty2003} employ consistency indices—GCI and CR—that are nearly identical.
Compared to methods \citep{aguaron2021,saaty2003},
Wu and Tu \citep{WuZhibin2021Mtac} changed more preferences, resulting in higher values of NPR and AOC while producing lower NV.
This is because the method in \citep{WuZhibin2021Mtac} considers both the acceptable inconsistency level and transitivity, making the modified PCM more reliable.
Due to the index-exchangeability condition imposing stricter constraints than transitivity, model (\ref{cp3poip11}) necessitates revising a greater number of preferences.
This finding indicates that to achieve more accurate and reliable ranking results, extremely coherent preferences from DMs are essential.

\begin{table}[H]
  \centering
  \caption{The simulation results by different inconsistency reduction methods}
\begin{tabular}{cllll}
\hline
\multicolumn{1}{l}{} & Methods     & AOC   & NPR   & NV     \\ \hline
\multirow{4}{*}{$n=4$} & Saaty \cite{saaty2003} & 1.54  & \textbf{1.00}  & 0.42   \\
                     & Aguar{\'o}n et al. \cite{aguaron2021} & 1.61  & 1.02  & 0.43   \\
                     & Wu and Tu \cite{WuZhibin2021Mtac}      & 2.37  & \textbf{1.00}  & 0.40   \\
                     & Model (\ref{cp3poip11})        & \textbf{0.75}  & 1.28  & \textbf{0.00}    \\\hline
\multirow{4}{*}{$n=5$} & Saaty \cite{saaty2003} & 1.59  & \textbf{1.06}  & 2.37   \\
                     & Aguar{\'o}n et al. \cite{aguaron2021} & \textbf{1.44}  & 1.07  & 2.51   \\
                     & Wu and Tu \cite{WuZhibin2021Mtac}      & 1.60  & 1.09  & 2.32   \\
                     & Model (\ref{cp3poip11})         & 2.76  & 2.48  & \textbf{0.00}   \\\hline
\multirow{4}{*}{$n=6$} & Saaty \cite{saaty2003} & 1.75  & 1.13  & 8.03   \\
                     & Aguar{\'o}n et al. \cite{aguaron2021} & \textbf{1.70}  & \textbf{1.07}  & 8.05   \\
                     & Wu and Tu \cite{WuZhibin2021Mtac}      & 2.25  & 1.24  & 7.95   \\
                     & Model (\ref{cp3poip11})         & 5.17  & 5.00  & \textbf{0.00}    \\\hline
\multirow{4}{*}{$n=7$} & Saaty \cite{saaty2003} & 2.01  & 1.30  & 24.90  \\
                     & Aguar{\'o}n et al. \cite{aguaron2021} & \textbf{1.80}  & \textbf{1.12}  & 25.02  \\
                     & Wu and Tu \cite{WuZhibin2021Mtac}      & 2.52  & 1.35  & 24.89  \\
                     & Model (\ref{cp3poip11})         & 8.99  & 9.24  & \textbf{0.00}    \\\hline
\multirow{4}{*}{$n=8$} & Saaty \cite{saaty2003} & \textbf{2.52}  & \textbf{1.50}  & 55.20  \\
                     & Aguar{\'o}n et al. \cite{aguaron2021} & 2.69  & 1.57  & 50.57  \\
                     & Wu and Tu \cite{WuZhibin2021Mtac}      & 3.08  & 1.64  & 48.42  \\
                     & Model (\ref{cp3poip11})         & 15.14 & 15.00 & \textbf{0.00}   \\\hline
\multirow{4}{*}{$n=9$} & Saaty \cite{saaty2003} & \textbf{2.88}  & \textbf{1.82}  & 110.35 \\
                     & Aguar{\'o}n et al. \cite{aguaron2021} & \textbf{2.88}  & 1.96  & 108.47 \\
                     & Wu and Tu \cite{WuZhibin2021Mtac}      & 3.24  & 2.01  & 104.28 \\
                     & Model (\ref{cp3poip11})        & 29.16 & 19.05 & \textbf{0.00}  \\\hline
\end{tabular}
\label{sl1000}
\end{table}

\section{Conclusions}\label{sec7}

In this paper, we introduce a procedure to deal with the violations of COP in AHP. 
If the PCM meets the index-exchangeability condition, the MNVDM method can derive the priority vector without violations.
However, if the PCM does not meet this condition, an optimization method designed to minimize information loss is employed to revise the contentious preferences. 
To ensure that practitioners can easily implement our models without requiring deep mathematical expertise, we have made the code open source and provide a step-by-step guide, which can be available at: \texttt{\href{https://github.com/Tommytutu/COP}{https://github.com/Tommytutu/COP}}.
The primary contributions of this paper are as follows:

\begin{itemize}
\setlength{\itemsep}{0pt}
\setlength{\parsep}{0pt}
\setlength{\parskip}{0pt}
\item[(1)] 
To determine whether a PCM satisfies the COP, we first analyze the relationships between transitivity, POP,  and POIP.
It concludes that  whether a PCM satisfies the COP depends on whether it meets the POIP conditions.
Subsequently, a necessary and sufficient condition (see Proposition \ref{theoremiff}) is proposed to detect the violations of COP.
This detection method is straightforward and independent of prioritization methods.

\item[(2)] Even when a satisfactory priority vector exists, as Bana e Costa and Vansnick \cite{BanaECostaCarlosA2008} demonstrated, methods like EM and LLSM may fail to derive a priority vector that meets COP. To address this issue, the MNVDM model, a two-stage optimization model where the first stage is to minimize the violations and the second stage can apply different deviation functions ${\cal{D}}(\mathbf{A}, \mathbf{W})$ to minimize the deviations, 
    is used to obtain the priority vector.  
    Simulation experiments and numerical examples show that the MNVDM model   derives the priority vector with the minimum NV.

\item[(3)] In some practical decision-making problems, the  preferences given by the DM may contain a high level of inconsistency or intransitivity. 
In such cases, no prioritization method can derive a coherent priority vector without violations.  
Consequently, we propose an optimization model designed to minimize information loss, thereby revising the preferences to comply with the COP.
 Rather than compelling the DM to accept the results generated by model  (\ref{cp3poip11}), we offer recommendations on how to amend their contradictory preferences.

\end{itemize}

The main drawback of this paper is the rapidly increasing complexity  of MNVDM and model (\ref{cp3poip11}) as the value of $n$ increases.
Therefore, we plan to propose more effective algorithms to solve model (\ref{cp3poip11}) in the future.
Incomplete PCMs have been prominent in preference modeling research and practice  \citep{szadoczki2023,csato2024choose}, however,
both the EM and LLSM can yield an ordering of the alternatives that contradicts the ordinally consistent preferences \citep{Faramondi2020}.
Hence, it would be valuable to investigate the COP in incomplete PCMs in future studies.
 Additionally, consensus achievement becomes an important area in decision analysis \citep{zhang2020overview}, we try to apply the proposed methods   to the consensus achievement of AHP group decision-making.

\bibliographystyle{plainnat} 
\bibliography{rankreversal}

\begin{thebibliography}{60}
\providecommand{\natexlab}[1]{#1}
\providecommand{\url}[1]{\texttt{#1}}
\expandafter\ifx\csname urlstyle\endcsname\relax
  \providecommand{\doi}[1]{doi: #1}\else
  \providecommand{\doi}{doi: \begingroup \urlstyle{rm}\Url}\fi

\bibitem[Aguar{\'{o}}n and Moreno-Jim{\'{e}}nez(2003)]{Aguarón2003}
Juan Aguar{\'{o}}n and Jos{\'{e}}~Mar{\'{i}}a Moreno-Jim{\'{e}}nez.
\newblock {The geometric consistency index: Approximated thresholds}.
\newblock \emph{European Journal of Operational Research}, 147\penalty0
  (1):\penalty0 137--145, 2003.
\newblock ISSN 03772217.

\bibitem[Aguar{\'o}n et~al.(2021)Aguar{\'o}n, Escobar, and
  Moreno-Jim{\'e}nez]{aguaron2021}
Juan Aguar{\'o}n, Mar{\'\i}a~Teresa Escobar, and Jos{\'e}~Mar{\'\i}a
  Moreno-Jim{\'e}nez.
\newblock Reducing inconsistency measured by the geometric consistency index in
  the analytic hierarchy process.
\newblock \emph{European Journal of Operational Research}, 288\penalty0
  (2):\penalty0 576--583, 2021.

\bibitem[{Bana e Costa} and Vansnick(2008)]{BanaECostaCarlosA2008}
Carlos~A. {Bana e Costa} and Jean~Claude Vansnick.
\newblock {A critical analysis of the eigenvalue method used to derive
  priorities in AHP}.
\newblock \emph{European Journal of Operational Research}, 187\penalty0
  (3):\penalty0 1422--1428, 2008.
\newblock ISSN 03772217.

\bibitem[Boz{\'o}ki and F{\"u}l{\"o}p(2018)]{bozoki2018}
S{\'a}ndor Boz{\'o}ki and J{\'a}nos F{\"u}l{\"o}p.
\newblock Efficient weight vectors from pairwise comparison matrices.
\newblock \emph{European Journal of Operational Research}, 264\penalty0
  (2):\penalty0 419--427, 2018.

\bibitem[Boz{\'{o}}ki et~al.(2015)Boz{\'{o}}ki, F{\"{u}}l{\"{o}}p, and
  Poesz]{Bozoki2015}
S{\'{a}}ndor Boz{\'{o}}ki, J{\'{a}}nos F{\"{u}}l{\"{o}}p, and Attila Poesz.
\newblock {On reducing inconsistency of pairwise comparison matrices below an
  acceptance threshold}.
\newblock \emph{Central European Journal of Operations Research}, 23\penalty0
  (4):\penalty0 849--866, 2015.
\newblock ISSN 16139178.

\bibitem[Brunelli(2015)]{brunelli2015}
Matteo Brunelli.
\newblock \emph{Introduction to the analytic hierarchy process}.
\newblock Springer, 2015.

\bibitem[Brunelli(2018)]{brunelli2018}
Matteo Brunelli.
\newblock A survey of inconsistency indices for pairwise comparisons.
\newblock \emph{International Journal of General Systems}, 47\penalty0
  (8):\penalty0 751--771, 2018.

\bibitem[Brunelli and Cavallo(2020{\natexlab{a}})]{Brunelli2020b}
Matteo Brunelli and Bice Cavallo.
\newblock {Incoherence measures and relations between coherence conditions for
  pairwise comparisons}.
\newblock \emph{Decisions in Economics and Finance}, 43\penalty0 (2):\penalty0
  613--635, 2020{\natexlab{a}}.
\newblock ISSN 11296569.

\bibitem[Brunelli and Cavallo(2020{\natexlab{b}})]{brunelli2020}
Matteo Brunelli and Bice Cavallo.
\newblock Distance-based measures of incoherence for pairwise comparisons.
\newblock \emph{Knowledge-Based Systems}, 187:\penalty0 104808,
  2020{\natexlab{b}}.

\bibitem[Brunelli and Fedrizzi(2024)]{brunelli2024}
Matteo Brunelli and Michele Fedrizzi.
\newblock Inconsistency indices for pairwise comparisons and the pareto
  dominance principle.
\newblock \emph{European Journal of Operational Research}, 312\penalty0
  (1):\penalty0 273--282, 2024.

\bibitem[Cavallo(2019)]{cavallo2019}
Bice Cavallo.
\newblock Coherent weights for pairwise comparison matrices and a mixed-integer
  linear programming problem.
\newblock \emph{Journal of Global Optimization}, 75\penalty0 (1):\penalty0
  143--161, 2019.

\bibitem[Cavallo and D'Apuzzo(2020)]{cavallo2020}
Bice Cavallo and Livia D'Apuzzo.
\newblock Preservation of preferences intensity of an inconsistent pairwise
  comparison matrix.
\newblock \emph{International Journal of Approximate Reasoning}, 116:\penalty0
  33--42, 2020.

\bibitem[Cavallo et~al.(2016)Cavallo, D'Apuzzo, and Basile]{cavallo2016}
Bice Cavallo, Livia D'Apuzzo, and Luciano Basile.
\newblock Weak consistency for ensuring priority vectors reliability.
\newblock \emph{Journal of Multi-Criteria Decision Analysis}, 23\penalty0
  (3-4):\penalty0 126--138, 2016.

\bibitem[Choo and Wedley(2004)]{CHOO2004}
Eng~Ung Choo and William~C Wedley.
\newblock A common framework for deriving preference values from pairwise
  comparison matrices.
\newblock \emph{Computers \& Operations Research}, 31\penalty0 (6):\penalty0
  893--908, 2004.

\bibitem[Crawford and Williams(1985)]{Crawford1985}
Gordon Crawford and Cindy Williams.
\newblock A note on the analysis of subjective judgment matrices.
\newblock \emph{Journal of Mathematical Psychology}, 29\penalty0 (4):\penalty0
  387--405, 1985.

\bibitem[Csat{\'o}(2019)]{csato2019}
L{\'a}szl{\'o} Csat{\'o}.
\newblock A characterization of the logarithmic least squares method.
\newblock \emph{European Journal of Operational Research}, 276\penalty0
  (1):\penalty0 212--216, 2019.

\bibitem[Csat{\'o}(2024{\natexlab{a}})]{csato2024}
L{\'a}szl{\'o} Csat{\'o}.
\newblock Right-left asymmetry of the eigenvector method: A simulation study.
\newblock \emph{European Journal of Operational Research}, 313\penalty0
  (2):\penalty0 708--717, 2024{\natexlab{a}}.

\bibitem[Csat{\'o}(2024{\natexlab{b}})]{csato2024choose}
L{\'a}szl{\'o} Csat{\'o}.
\newblock How to choose a completion method for pairwise comparison matrices
  with missing entries: An axiomatic result.
\newblock \emph{International Journal of Approximate Reasoning}, 164:\penalty0
  109063, 2024{\natexlab{b}}.

\bibitem[Csat{\'o} and Petr{\'o}czy(2021)]{csato2021}
L{\'a}szl{\'o} Csat{\'o} and D{\'o}ra~Gr{\'e}ta Petr{\'o}czy.
\newblock On the monotonicity of the eigenvector method.
\newblock \emph{European Journal of Operational Research}, 292\penalty0
  (1):\penalty0 230--237, 2021.

\bibitem[Faramondi et~al.(2020)Faramondi, Oliva, and Boz{\'o}ki]{Faramondi2020}
Luca Faramondi, Gabriele Oliva, and S{\'a}ndor Boz{\'o}ki.
\newblock Incomplete analytic hierarchy process with minimum weighted ordinal
  violations.
\newblock \emph{International Journal of General Systems}, 49\penalty0
  (6):\penalty0 574--601, 2020.

\bibitem[Genest and Rivest(1994)]{genest1994}
Christian Genest and Louis~Paul Rivest.
\newblock A statistical look at saaty's method of estimating pairwise
  preferences expressed on a ratio scale.
\newblock \emph{Journal of Mathematical Psychology}, 38\penalty0 (4):\penalty0
  477--496, 1994.

\bibitem[Golany and Kress(1993)]{golany1993}
Boaz Golany and Moshe Kress.
\newblock A multicriteria evaluation of methods for obtaining weights from
  ratio-scale matrices.
\newblock \emph{European Journal of Operational Research}, 69\penalty0
  (2):\penalty0 210--220, 1993.

\bibitem[Grzybowski(2016)]{grzybowski2016new}
Andrzej~Z Grzybowski.
\newblock New results on inconsistency indices and their relationship with the
  quality of priority vector estimation.
\newblock \emph{Expert Systems with Applications}, 43:\penalty0 197--212, 2016.

\bibitem[Johnson et~al.(1979)Johnson, Beine, and Wang]{Johnson1979}
Charles~R. Johnson, William~B. Beine, and Theodore~J. Wang.
\newblock {Right-left asymmetry in an eigenvector ranking procedure}.
\newblock \emph{Journal of Mathematical Psychology}, 19\penalty0 (1):\penalty0
  61--64, 1979.
\newblock ISSN 10960880.
\newblock \doi{10.1016/0022-2496(79)90005-1}.

\bibitem[Karapetrovic and Rosenbloom(1999)]{karapetrovic1999}
Stanislav Karapetrovic and ES~Rosenbloom.
\newblock {A quality control approach to consistency paradoxes in AHP}.
\newblock \emph{European Journal of Operational Research}, 119\penalty0
  (3):\penalty0 704--718, 1999.

\bibitem[Kazibudzki(2021)]{kazibudzki2021}
P.~Kazibudzki.
\newblock On the statistical discrepancy and affinity of priority vector
  heuristics in pairwise-comparison-based methods.
\newblock \emph{Entropy}, 23\penalty0 (9):\penalty0 1150, 2021.

\bibitem[Kazibudzki(2019{\natexlab{a}})]{kazibudzki2019}
Paul~Thaddeus Kazibudzki.
\newblock An examination of ranking quality for simulated pairwise judgments in
  relation to performance of the selected consistency measure.
\newblock \emph{Advances in Operations Research}, 2019, 2019{\natexlab{a}}.

\bibitem[Kazibudzki(2019{\natexlab{b}})]{kazibudzki2019q}
Paul~Thaddeus Kazibudzki.
\newblock The quality of ranking during simulated pairwise judgments for
  examined approximation procedures.
\newblock \emph{Modelling and Simulation in Engineering}, 2019,
  2019{\natexlab{b}}.

\bibitem[Kazibudzki(2022)]{kazibudzki2022}
Pawe{\l}~Tadeusz Kazibudzki.
\newblock On estimation of priority vectors derived from inconsistent pairwise
  comparison matrices.
\newblock \emph{Journal of Applied Mathematics and Computational Mechanics},
  21\penalty0 (4):\penalty0 52--59, 2022.

\bibitem[Koczkodaj and Urban(2018)]{koczkodaj2018}
Waldemar~W Koczkodaj and Roman Urban.
\newblock Axiomatization of inconsistency indicators for pairwise comparisons.
\newblock \emph{International Journal of Approximate Reasoning}, 94:\penalty0
  18--29, 2018.

\bibitem[Ku{\l}akowski(2015)]{kulakowski2015}
Konrad Ku{\l}akowski.
\newblock {Notes on order preservation and consistency in AHP}.
\newblock \emph{European Journal of Operational Research}, 245\penalty0
  (1):\penalty0 333--337, 2015.

\bibitem[Ku{\l}akowski et~al.(2019)Ku{\l}akowski, Mazurek, Ram{\'\i}k, and
  Soltys]{kulakowski2019}
Konrad Ku{\l}akowski, Ji{\v{r}}{\'\i} Mazurek, Jaroslav Ram{\'\i}k, and Michael
  Soltys.
\newblock When is the condition of order preservation met?
\newblock \emph{European Journal of Operational Research}, 277\penalty0
  (1):\penalty0 248--254, 2019.

\bibitem[Li et~al.(2019)Li, Dong, Xu, Chiclana, Herrera-Viedma, and
  Herrera]{li2019overview}
Cong~Cong Li, Yu~Cheng Dong, Ye~Jun Xu, Francisco Chiclana, Enrique
  Herrera-Viedma, and Francisco Herrera.
\newblock An overview on managing additive consistency of reciprocal preference
  relations for consistency-driven decision making and fusion: Taxonomy and
  future directions.
\newblock \emph{Information Fusion}, 52:\penalty0 143--156, 2019.

\bibitem[Li et~al.(2022)Li, Zhang, and Yu]{li2022consensus}
Zhuo~Lin Li, Zhen Zhang, and Wen~Yu Yu.
\newblock Consensus reaching with consistency control in group decision making
  with incomplete hesitant fuzzy linguistic preference relations.
\newblock \emph{Computers \& Industrial Engineering}, 170:\penalty0 108311,
  2022.

\bibitem[Liu et~al.(2021)Liu, Zou, and Pedrycz]{Liu2021}
Fang Liu, Shucai Zou, and Witold Pedrycz.
\newblock Measuring weak consistency and weak transitivity of pairwise
  comparison matrices.
\newblock \emph{IEEE Transactions on Cybernetics}, 53\penalty0 (1):\penalty0
  303--314, 2021.

\bibitem[Mazurek(2022)]{mazurek2022}
Ji{\v{r}}{\'\i} Mazurek.
\newblock New preference violation indices for the condition of order
  preservation.
\newblock \emph{RAIRO-Operations Research}, 56\penalty0 (1):\penalty0 367--380,
  2022.

\bibitem[Mazurek and Ku{\l}akowski(2020)]{JiriMazurek2020}
Ji{\v{r}}{\'\i} Mazurek and Konrad Ku{\l}akowski.
\newblock Satisfaction of the condition of order preservation : a simulation
  study.
\newblock \emph{Operations Research and Decisions}, 30\penalty0 (2):\penalty0
  77--89, 2020.
\newblock ISSN 23916060.

\bibitem[Mazurek and Ram{\'\i}k(2019)]{Mazurek2019}
Ji{\v{r}}{\'\i} Mazurek and Jaroslav Ram{\'\i}k.
\newblock {Some new properties of inconsistent pairwise comparisons matrices}.
\newblock \emph{International Journal of Approximate Reasoning}, 113:\penalty0
  119--132, 2019.
\newblock ISSN 0888613X.

\bibitem[Miller(1956)]{miller1956}
George~A Miller.
\newblock The magical number seven, plus or minus two: Some limits on our
  capacity for processing information.
\newblock \emph{Psychological Review}, 63\penalty0 (2):\penalty0 81, 1956.

\bibitem[{P. T. Kazibudzki}(2016)]{kazibudzki2016}
{P. T. Kazibudzki}.
\newblock An examination of performance relations among selected consistency
  measures for simulated pairwise judgments.
\newblock \emph{Annals of Operations Research}, 244\penalty0 (2):\penalty0
  525--544, 2016.

\bibitem[{P. T. Kazibudzki}(2023)]{kazibudzki2023}
{P. T. Kazibudzki}.
\newblock The uncertainty related to the inexactitude of prioritization based
  on consistent pairwise comparisons.
\newblock \emph{Plos One}, 18\penalty0 (9):\penalty0 1--30, 09 2023.

\bibitem[Pant et~al.(2022)Pant, Kumar, Ram, Klochkov, and Sharma]{pant2022}
Sangeeta Pant, Anuj Kumar, Mangey Ram, Yury Klochkov, and Hitesh~Kumar Sharma.
\newblock Consistency indices in analytic hierarchy process: a review.
\newblock \emph{Mathematics}, 10\penalty0 (8):\penalty0 1206, 2022.

\bibitem[Regenwetter et~al.(2011)Regenwetter, Dana, and
  Davis-Stober]{regenwetter2011}
Michel Regenwetter, Jason Dana, and Clintin~P Davis-Stober.
\newblock Transitivity of preferences.
\newblock \emph{Psychological review}, 118\penalty0 (1):\penalty0 42, 2011.

\bibitem[Saaty(1977)]{saaty1977}
Thomas~L Saaty.
\newblock A scaling method for priorities in hierarchical structures.
\newblock \emph{Journal of Mathematical Psychology}, 15\penalty0 (3):\penalty0
  234--281, 1977.

\bibitem[Saaty(1980)]{Saaty1980}
Thomas~L Saaty.
\newblock \emph{The Analytic Hierarchy Process}.
\newblock McGraw-Hill, New York, 1980.

\bibitem[Saaty(2003)]{saaty2003}
Thomas~L Saaty.
\newblock {Decision-making with the {AHP}: Why is the principal eigenvector
  necessary}.
\newblock \emph{European Journal of Operational Research}, 145\penalty0
  (1):\penalty0 85--91, 2003.

\bibitem[Siraj et~al.(2012)Siraj, Mikhailov, and Keane]{siraj2012}
Sajid Siraj, Ludmil Mikhailov, and John Keane.
\newblock A heuristic method to rectify intransitive judgments in pairwise
  comparison matrices.
\newblock \emph{European Journal of Operational Research}, 216\penalty0
  (2):\penalty0 420--428, 2012.

\bibitem[Sz{\'a}doczki et~al.(2023)Sz{\'a}doczki, Boz{\'o}ki, Juh{\'a}sz,
  Kadenko, and Tsyganok]{szadoczki2023}
Zsombor Sz{\'a}doczki, S{\'a}ndor Boz{\'o}ki, Patrik Juh{\'a}sz, Sergii~V
  Kadenko, and Vitaliy Tsyganok.
\newblock Incomplete pairwise comparison matrices based on graphs with average
  degree approximately 3.
\newblock \emph{Annals of Operations Research}, 326\penalty0 (2):\penalty0
  783--807, 2023.

\bibitem[Tu and Wu(2022)]{Tuwu2022}
Jian~Cheng Tu and Zhi~Bin Wu.
\newblock H-rank consensus models for fuzzy preference relations considering
  eliminating rank violations.
\newblock \emph{IEEE Transactions on Fuzzy Systems}, 30\penalty0 (6):\penalty0
  2004--2018, 2022.
\newblock \doi{10.1109/TFUZZ.2021.3073238}.

\bibitem[Tu and Wu(2023)]{tu2023}
Jian~Cheng Tu and ZhiBin Wu.
\newblock Analytic hierarchy process rank reversals: causes and solutions.
\newblock \emph{Annals of Operations Research}, pages 1--25, 2023.

\bibitem[Tversky(1969)]{tversky1969}
Amos Tversky.
\newblock Intransitivity of preferences.
\newblock \emph{Psychological review}, 76\penalty0 (1):\penalty0 31, 1969.

\bibitem[Wang et~al.(2021{\natexlab{a}})Wang, Kou, and Peng]{wang2021b}
Hao~Min Wang, Gang Kou, and Yi~Peng.
\newblock An iterative algorithm to derive priority from large-scale sparse
  pairwise comparison matrix.
\newblock \emph{IEEE Transactions on Systems, Man, and Cybernetics: Systems},
  52\penalty0 (5):\penalty0 3038--3051, 2021{\natexlab{a}}.

\bibitem[Wang et~al.(2021{\natexlab{b}})Wang, Peng, and Kou]{wang2021}
Hao~Min Wang, Yi~Peng, and Gang Kou.
\newblock A two-stage ranking method to minimize ordinal violation for pairwise
  comparisons.
\newblock \emph{Applied Soft Computing}, 106:\penalty0 107287,
  2021{\natexlab{b}}.

\bibitem[Wu and Tu(2021)]{WuZhibin2021Mtac}
Zhi~Bin Wu and Jian~Cheng Tu.
\newblock {Managing transitivity and consistency of preferences in AHP group
  decision making based on minimum modifications}.
\newblock \emph{Information Fusion}, 67:\penalty0 125--135, 2021.

\bibitem[Xu et~al.(2021)Xu, Li, Cabrerizo, Chiclana, and
  Herrera-Viedma]{Xu2021}
Ye~Jun Xu, Meng~Qi Li, Francisco~Javier Cabrerizo, Francisco Chiclana, and
  Enrique Herrera-Viedma.
\newblock Algorithms to detect and rectify multiplicative and ordinal
  inconsistencies of fuzzy preference relations.
\newblock \emph{IEEE Transactions on Systems, Man, and Cybernetics: Systems},
  51\penalty0 (6):\penalty0 3498--3511, 2021.
\newblock ISSN 21682232.

\bibitem[Xu et~al.(2023)Xu, Wang, Chiclana, and Herrera-Viedma]{xu2023local}
Ye~Jun Xu, Qian~Qian Wang, Francisco Chiclana, and Enrique Herrera-Viedma.
\newblock A local adjustment method to improve multiplicative consistency of
  fuzzy reciprocal preference relations.
\newblock \emph{IEEE Transactions on Systems, Man, and Cybernetics: Systems},
  53\penalty0 (9):\penalty0 5702--5714, 2023.

\bibitem[Yuen(2024)]{yuen2024}
Kevin Kam~Fung Yuen.
\newblock Inverse gram matrix methods for prioritization in analytic hierarchy
  process: Explainability of weighted least squares optimization method.
\newblock \emph{arXiv preprint arXiv:2401.01190}, 2024.

\bibitem[Zhang et~al.(2020)Zhang, Zhao, Kou, Li, Dong, and
  Herrera]{zhang2020overview}
Heng~Jie Zhang, Si~Hai Zhao, Gang Kou, Cong~Cong Li, Yu~Cheng Dong, and
  Francisco Herrera.
\newblock An overview on feedback mechanisms with minimum adjustment or cost in
  consensus reaching in group decision making: Research paradigms and
  challenges.
\newblock \emph{Information Fusion}, 60:\penalty0 65--79, 2020.

\bibitem[Zhang et~al.(2021{\natexlab{a}})Zhang, Kou, Peng, and
  Zhang]{Zhang2021}
Jiulong Zhang, Gang Kou, Yi~Peng, and Yu~Zhang.
\newblock Estimating priorities from relative deviations in pairwise comparison
  matrices.
\newblock \emph{Information Sciences}, 552:\penalty0 310--327,
  2021{\natexlab{a}}.

\bibitem[Zhang et~al.(2021{\natexlab{b}})Zhang, Kou, Yu, and
  Gao]{zhang2021consistency}
Zhen Zhang, Xin~Yue Kou, Wen~Yu Yu, and Yuan Gao.
\newblock Consistency improvement for fuzzy preference relations with
  self-confidence: An application in two-sided matching decision making.
\newblock \emph{Journal of the Operational Research Society}, 72\penalty0
  (8):\penalty0 1914--1927, 2021{\natexlab{b}}.

\end{thebibliography}

\section*{Online Supplement}
\label{sec:appendix}

\subsection{Supplement for Section \ref{22}}
\label{subsec:1}
(1) \emph{Model by Brunelli and Cavallo \cite{Brunelli2020b}}

Brunelli and Cavallo \cite{Brunelli2020b}  suggested the following optimization model for obtaining the closest PCM that satisfies  the index-exchangeability condition:
\begin{equation}
\label{Bru2020}
  \begin{array}{l}
  \min \quad \mathbf{I}_{\mathcal{I} \mathcal{E}}^{\varepsilon}(\mathbf{A})=\frac{2}{n(n-1)} \sum\limits_{i= 1}^{n-1}\sum\limits_{j=i+1}^{n} ( d_{i j}^{+}+d_{i j}^{-})\\
  s.t.
    \left\{ \begin{array}{ll}
    x_{i j}-\log(a_{i j})=d_{i j}^{+}-d_{i j}^{-} & i, j \in N\\
d_{i j}^{+}, d_{i j}^{-} \geq 0  & i, j \in N\\
x_{i j}+x_{j i}=0 & i, j \in N \\
x_{i k}-x_{r s} \leq M y_{i k r s} & i, k, r, s \in N \\
x_{i r}-x_{k s} \geq \varepsilon-M\left(1-y_{i k r s}\right) & i, k, r, s \in N \\
y_{i k r s} \in\{0,1\} & i, k, r, s \in N
  \end{array} \right.
  \end{array}
\end{equation}

For model (\ref{Bru2020}), $x_{i j}$, $d_{i j}^{+}$, $d_{i j}^{-}$,  and $y_{i k r s}$ are decision variables,  $M$ is a big value and $\varepsilon$ is a positive number close to zero.
The first two constraints are used to calculate the logarithmic Mahaltonian distance between original PCM $\mathbf{A}=(a_{ij})_{n\times n} \subset X \times X$ and the revised PCM $\mathbf{\overline{A}}=(\overline{{a}_{ij}})_{n\times n} \subset X \times X$ where  $x_{ij}=\log \overline{{a}_{ij}}$;
the third constraint ensures that the revised PCM  $\mathbf{\overline{A}}$ satisfies the reciprocality;
the last three constraints show the relationship between the binary variable  $y_{i k r s}$ and the decision variables ($x_{i k} $ and  $x_{r s} $).
For more information about model (\ref{Bru2020}), please refer to model (26) in \cite{Brunelli2020b}.

(2) \emph{The prioritization method by Mazurek and Ram{\'\i}k \cite{Mazurek2019}}

Mazurek and Ram{\'\i}k \cite{Mazurek2019} proposed the following prioritization method  considering the POIP condition:

\begin{equation}
\label{cp3ipops21}
  \begin{array}{l}

{\cal D}({\bf{A}},{\bf{W}})=\min \left\{\max \left\{\frac{a_{i j} w_{j}}{w_{i}}, \frac{w_{i}}{a_{i j} w_{j}}\right\} \mid i, j \in\{1, \ldots, n\}\right\}\\
  s.t.
    \left\{ \begin{array}{ll}
    \sum_{i=1}^{n} w_{i}=1, w_{i} \geq  \varepsilon  \quad \forall i \in N\\
    \frac{w_{i}}{w_{j}}-\frac{w_{k}}{w_{l}} \geq \varepsilon, \quad \forall(i,j,k,l) \in I^{(4)}(\mathbf{A})\\
  \end{array} \right.
  \end{array}
\end{equation}
where $I^{(4)}(\mathbf{A})=\{(i, j, k, l) \mid i, j, k, l \in\{1, \ldots, n\}, a_{i j}>1, a_{k l}>1, a_{i j}>a_{k l}\}$.

Although model (\ref{cp3ipops21}) derives the priority vector considering the POIP conditions, it has two shortcomings that need improvement.
Firstly, the feasible solution set of model (\ref{cp3ipops21}) becomes empty when  $\bf{A}$ violates the index-exchangeability condition.
Secondly, model (\ref{cp3ipops21}) is a non-convex optimization model, hence, it may be difficult to get the optimal solution.
Since Mazurek and Ram{\'\i}k \cite{Mazurek2019} did not provide the solution process of  model (\ref{cp3ipops21}), we utilized the  \texttt{fmincon} function in  \texttt{Matlab} to solve  model (\ref{cp3ipops21}), with all parameters of the \texttt{fmincon} function set to their default values.

\subsection{Supplement for Section \ref{sec3}}
(1) \emph{The inconsistency index   by Ku{\l}akowski et al. \cite{kulakowski2019}}

The minimal error index (MEI) is defined as:
\begin{equation}
\text{MEI}(\mathbf{A}) = \max _{i, j}\left\{a_{i j} \frac{w_{j}}{w_{i}}-1\right\}
\end{equation}
where the priority vector $\mathbf{w}$  can be derived by the existing methods such as EM and LLSM. Ku{\l}akowski et al. \cite{kulakowski2019} suggested that $\mathbf{w}$  can also be obtained by solving the following optimization model:
\begin{equation}\label{modelE}
\left\{\begin{array}{l}
\min \max\limits_{i, j}\left\{a_{i j} \frac{w_{j}}{w_{i}}-1\right\}\\
\text { s.t. } \sum\limits_{i=1}^{n} w_{i}=1, w_{i} > 0
\end{array}\right.
\end{equation}

(2) \emph{The additive relative deviation interconnection (ARDI) model  by Zhang et al. \cite{Zhang2021}}

\begin{equation}
\begin{array}{lll}
\min\limits_{\boldsymbol{w}, \boldsymbol{\varepsilon}, \boldsymbol{v}} & \|\boldsymbol{v}\|_{\hat{p}_{1}} & \\
\text { s.t. } & a_{i j}=\frac{w_{i} \cdot \varepsilon_{i j}^{i}}{w_{j}\cdot \varepsilon_{i j}^{i}}, \quad \forall i, j \in N: i<j, \\
& \left\|\left(\varepsilon_{i j}^{i}, \varepsilon_{i j}^{j}\right)\right\|_{q_{1}} \leqslant v_{i j}, \quad \forall i, j \in N: i<j, \\
& \boldsymbol{e}^{\top} \boldsymbol{w}=1, & \\
& \boldsymbol{w}>\mathbf{0} . &
\end{array}
\label{ARDIm}
\end{equation}

$\boldsymbol{v}=\left(v_{i j}\right)$  for all  $i, j \in N$ , and  $\hat{p}_{1}, q_{1} \geqslant 1 $ are real numbers.
In this paper, we set $\hat{p}_{1}=1$ and $q_{1} =1 $.
Then model (\ref{ARDIm}) is equivalent to the following linear programming model:

\begin{equation}
\label{ARDIlp}
  \begin{array}{l}
  \min \quad  \sum\limits_{i= 1}^{n-1}\sum\limits_{j=i+1}^{n} v_{ij}\\
  s.t.
    \left\{ \begin{array}{ll}
a_{i j}=\frac{w_{i}+\varepsilon_{i j}^{i}}{w_{j}+\varepsilon_{i j}^{i}} & \forall i, j \in N: i<j \\
|\varepsilon_{i j}^{i}| + |\varepsilon_{i j}^{j}| \leq  v_{ij} & \forall i, j \in N: i<j \\
\sum\limits_{i=1}^{n} w_{i}=1 & \\
w_{i} > 0 & \forall i \in N\\
  \end{array} \right.
  \end{array}
\end{equation}

\subsection{Supplement for Section \ref{sec4}}
(1) \emph{The proof of Proposition \ref{theoremiff}}

Necessity: see the proof of Proposition 4.2 in  Cavallo and D'Apuzzo \cite{cavallo2020}.

Sufficiency:  if $\mathbf{A} $ satisfies the index-exchangeability condition, that is,
$a_{ij}> a_{kl}  \Leftrightarrow a_{ik}> a_{jl}, i,j,k,l\in N$,
then there exists a priority vector $\textbf{w} = ({w_1},{w_2},\ldots,{w_n})^T$ where $\frac{w_i}{w_j}>\frac{w_k}{w_l}$.
Because $\frac{w_i}{w_j}>\frac{w_k}{w_l}\Leftrightarrow \frac{w_i}{w_k}>\frac{w_j}{w_l}$, we can obtain
$a_{ij}> a_{kl} \Leftrightarrow \frac{w_i}{w_j}>\frac{w_k}{w_l}$ and  $a_{ik}> a_{jl}\Leftrightarrow \frac{w_i}{w_k}>\frac{w_j}{w_l}$.
The proof of $a_{ij}=a_{kl}  \Leftrightarrow a_{ik}= a_{jl}, i,j,k,l\in N$ is similar.
Hence, $\textbf{w}$ satisfies the POIP conditions.

Thus,  the  index-exchangeability condition can be considered as the  necessary and sufficient  condition to judge whether there exists a priority vector meeting the POIP.

(2) \emph{The 15 cases in Example \ref{eample11}}

For $\mathbf{A}$ in Example \ref{eample11}, we need to consider the following 15 cases:

\begin{itemize}
\setlength{\itemsep}{0pt}
\setlength{\parsep}{0pt}
\setlength{\parskip}{0pt}
\item[1)] $a_{12}=2<a_{13}=4$ and $a_{11}=1<a_{23}=3$;
\item[2)] $a_{12}=2<a_{14}=9$ and $a_{11}=1<a_{24}=7$;
\item[3)] $a_{12}=2<a_{23}=3$ and $a_{12}=2<a_{23}=3$;
\item[4)] $a_{12}=2<a_{24}=7$ and $a_{12}=2<a_{24}=7$;
\item[5)] $a_{12}=2<a_{34}=5$ and $a_{13}=4<a_{24}=7$;
\item[6)] $a_{13}=4<a_{14}=9$ and $a_{11}=1<a_{34}=5$;
\item[7)] $a_{13}=4>a_{23}=3$ and $a_{12}=2>a_{33}=1$;
\item[8)] $a_{13}=4<a_{24}=7$ and $a_{12}=2<a_{34}=5$;
\item[9)] $a_{13}=4<a_{34}=5$ and $a_{13}=4<a_{34}=5$;
\item[10)] $a_{14}=9>a_{23}=3$ and $a_{12}=2>a_{43}=1/5$;
\item[11)] $a_{14}=9>a_{24}=7$ and $a_{12}=2>a_{44}=1$;
\item[12)] $a_{14}=9>a_{34}=5$ and $a_{13}=4>a_{44}=1$;
\item[13)] $a_{23}=3<a_{24}=7$ and $a_{22}=1<a_{34}=5$;
\item[14)] $a_{23}=3<a_{34}=5$ and $a_{23}=3<a_{34}=5$;
\item[15)] $a_{24}=7>a_{34}=5$ and $a_{23}=3>a_{44}=1$.
\end{itemize}

\subsection{Supplement for Section \ref{sec5}}

(1) \emph{The detailed calculation process of NV}

By combining (\ref{NR1}), (\ref{lea22}) and (\ref{wwaa}),  NV can be computed by the following three cases:

\begin{itemize}
\setlength{\itemsep}{0pt}
\setlength{\parsep}{0pt}
\setlength{\parskip}{0pt}
\item[1)] if $a_{ij}> a_{kl}$ but $w_i/w_j<w_k/w_l$:
\begin{equation}
\text{NV}_1=\sum\limits_{i= 1}^n\sum\limits_{j=i+1}^n\sum\limits_{k=1}^n\sum\limits_{l=k+1}^n
l_{ijkl}^a(1-e_{ijkl}^w-l_{ijkl}^w).
\end{equation}

\item[2)] if $a_{ij}> a_{kl}$ but $w_i/w_j= w_k/w_l$:
\begin{equation}
\text{NV}_2=\sum\limits_{i= 1}^n\sum\limits_{j=i+1}^n\sum\limits_{k=1}^n\sum\limits_{l=k+1}^n
\frac{1}{2}l_{ijkl}^ae_{ijkl}^w.
\end{equation}

\item[3)] if $a_{ij}= a_{kl}$ but $w_i/w_j \neq w_k/w_l$:
\begin{equation}
\text{NV}_3=\sum\limits_{i= 1}^n\sum\limits_{j=i+1}^n\sum\limits_{k=1}^n\sum\limits_{l=k+1}^n
\frac{1}{2}e_{ijkl}^a(1-e_{ijkl}^w).
\end{equation}

\end{itemize}

Thus,
\begin{equation} 
\begin{split}
\text{NV}= & \text{NV}_1 +\text{NV}_2 +\text{NV}_3\\
  = & \sum\limits_{i= 1}^n\sum\limits_{j=i+1}^n\sum\limits_{k=1}^n\sum\limits_{l=k+1}^n
      (l_{ijkl}^a(1-l_{ijkl}^w-\frac{1}{2}e_{ijkl}^w)\\
    & + \frac{1}{2}e_{ijkl}^a(1-e_{ijkl}^w)).\\
\end{split}
\end{equation}

(2) \emph{The proof of Proposition \ref{propmlp}}

Let $y_i=\log w_i$. $y_i$ and $w_i$ can be mutually transformed by
\begin{equation}
w_i=\frac{\exp(y_i)}{\sum\limits_{j= 1}^n \exp(y_j)}, \quad i \in N.
\end{equation}
Then model (\ref{cp3poip}) can be converted into the following  mixed-integer linear model (\ref{cp33poip}) for all $\forall i, k \in N, j>i, l>k$:

\begin{equation}
\label{cp33poip}
  \begin{array}{l}
   \min  \quad \text{NV}= \sum\limits_{i= 1}^{n-1}\sum\limits_{j=i+1}^n\sum\limits_{k=1}^{n-1}\sum\limits_{l=k+1}^n
      (l_{ijkl}^a(1-l_{ijkl}^w-\frac{1}{2}e_{ijkl}^w)\\
  \quad\quad\quad\quad\quad\quad   + \frac{1}{2}e_{ijkl}^a(1-e_{ijkl}^w))\\
  s.t.
    \left\{ \begin{array}{lll}
    -y_i-y_l+ (y_k+y_j)- M(1 - l_{ijkl}^w) \\+\varepsilon \leq 0,  & (\ref{cp33poip}-1)\\
    y_i+y_l- (y_k+y_j)- Ml_{ijkl}^w \leq0,  & (\ref{cp33poip}-2)\\
     -y_i-y_l+ (y_k+y_j)- M(1 - e_{ijkl}^w) \leq 0,  & (\ref{cp33poip}-3)\\
     y_i+y_l- (y_k+y_j)- M(1 - e_{ijkl}^w) \leq 0, & (\ref{cp33poip}-4)\\
     y_i+y_l- (y_k+y_j)- M(l_{ijkl}^w+e_{ijkl}^w)\\  +\varepsilon \leq 0,  & (\ref{cp33poip}-5)\\
     l_{ijkl}^w+e_{ijkl}^w \leq 1,& (\ref{cp33poip}-6)\\
     \sum_{i} y_{i}=0,  & (\ref{cp33poip}-7)\\
     l_{ijkl}^w, e_{ijkl}^w \in \{0,1\}, & (\ref{cp33poip}-8)\\
  \end{array} \right.
  \end{array}
\end{equation}
where $e_{ijkl}^w$, $e_{ijkl}^w$ and $y_i$ are decision variables, $M$ is a big value and $\varepsilon$ is a positive number close to zero.

(3) \emph{MNVEM}

The EM can be mathematically formulated as a convex optimization model, which is detailed in model (25) in reference  \citep{Bozoki2015}.
 To design model  (\ref{cp3ipops2common}) using the EM, we can make the following modifications:

 \begin{equation}
\label{cp3ipops2EM}
  \begin{array}{l}
  \min \quad \lambda\\
  s.t.
    \left\{ \begin{array}{ll}
    \sum\limits_{j= 1}^{n} \exp({{a}_{i j}+y_{j}-y_{i}}) \leq \lambda, i=1, \ldots, n\\
    \text{NV}= \text{NV}^*\\
    \text{constraints (\ref{cp33poip}-1)-(\ref{cp33poip}-8)} \\
  \end{array} \right.
  \end{array}
\end{equation}
However, it's important to note that incorporating the EM approach in model (\ref{cp3ipops2EM}) may introduce additional complexity due to the presence of exponential functions in the constraints. This makes model  (\ref{cp3ipops2EM}) more challenging to solve compared to model (\ref{cp3ipops2}).

(4) \emph{MNVMEM}

The two-stage model to  minimize the number of violations based on the model MEM by Ku{\l}akowski et al. \cite{kulakowski2019} (MNVMEM) is designed as follows:

\begin{equation}
\label{modelNVE}
  \begin{array}{l}
  \min \max\limits_{i, j}\left\{a_{i j} \frac{w_{j}}{w_{i}}-1\right\}\\
  s.t.
    \left\{ \begin{array}{ll}
    \text{NV}= \text{NV}^*\\
\text{constraints (\ref{cp33poip}-1)-(\ref{cp33poip}-8)} \\
  \end{array} \right.
  \end{array}
\end{equation}

(5) \emph{MNVLSDM}

The MNVLSDM, the two-stage model to  minimize the number of violations based on the LSDM \cite{kazibudzki2022,kazibudzki2021}, is designed as follows:
\begin{equation}
\label{modelNVLSD1}
  \begin{array}{l}
  \min \quad {\cal{D}}(\mathbf{A}, \mathbf{W})=\sum\limits_{i=1}^{n} \ln ^{2}\left(\sum\limits_{j=1}^{n}\left(a_{i j} w_{j} / (n w_{i})\right)\right)\\
  s.t.
    \left\{ \begin{array}{ll}
    \text{NV}= \text{NV}^*\\
\text{constraints (\ref{cp3poip}-1)-(\ref{cp3poip}-8)}
  \end{array} \right.
  \end{array}
\end{equation}

In model (\ref{modelNVLSD1}), the constraints include fractional functions. 
Thus, the relaxation form of model (\ref{modelNVLSD1}) is a non-convex problem which is difficult to find the optimal solution. 
To address this, we introduce the substitution  ($w_i = \exp(y_i)$), transforming model (\ref{modelNVLSD1}) into the following equivalent form:

\begin{equation}
\label{modelNVLSD}
  \begin{array}{l}
  \min \quad {\cal{D}}(\mathbf{A}, \mathbf{W})=\sum\limits_{i=1}^{n} \ln ^{2}\left(\sum\limits_{j=1}^{n}\left(a_{i j}\exp(y_j -y_i)/ n )\right)\right)\\
  s.t.
    \left\{ \begin{array}{ll}
    \text{NV}= \text{NV}^*\\
\text{constraints (\ref{cp33poip}-1)-(\ref{cp33poip}-8)}
  \end{array} \right.
  \end{array}
\end{equation}

Given that both the logarithmic and exponential functions in the objective function of model (\ref{modelNVLSD}) are convex functions, 
model (\ref{modelNVLSD}) becomes a mixed-integer convex programming model. This can be solved by most  optimization solvers, such as  \texttt{Ipopt}.

(6) \emph{MNVARDI}

The MNVARDI, the two-stage model to  minimize the number of violations based on the ARDI \cite{Zhang2021}, is designed as follows:

\begin{equation}
\label{modelNVARDI1}
  \begin{array}{l}
  \min \quad \quad {\cal{D}}(\mathbf{A}, \mathbf{W})= \sum\limits_{i= 1}^{n-1}\sum\limits_{j=i+1}^{n} v_{ij}\\
  s.t.
    \left\{ \begin{array}{ll}
a_{i j}=\frac{w_{i}+\varepsilon_{i j}^{i}}{w_{j}+\varepsilon_{i j}^{j}} & \forall i, j \in N: i<j \\
|\varepsilon_{i j}^{i}| + |\varepsilon_{i j}^{j}| \leq  v_{ij} & \forall i, j \in N: i<j \\
\text{NV}= \text{NV}^*\\
\text{constraints (\ref{cp3poip}-1)-(\ref{cp3poip}-8)}\\
  \end{array} \right.
  \end{array}
\end{equation}

In model (\ref{modelNVARDI1}), there are  fractional functions in the constraints. Thus, the relaxation form of model (\ref{modelNVARDI1}) is a non convex problem which is difficult to find the optimal solution. 
To solve this model, let  $w_i = \exp(y_i)$, then model (\ref{modelNVARDI1}) has the following equivalent form:

\begin{equation}
\label{modelNVARDI}
  \begin{array}{l}
  \min \quad \quad {\cal{D}}(\mathbf{A}, \mathbf{W})= \sum\limits_{i= 1}^{n-1}\sum\limits_{j=i+1}^{n} v_{ij}\\
  s.t.
    \left\{ \begin{array}{ll}
|a_{i j} \exp(y_j) - \exp(y_i)| \leq |\varepsilon_{i j}^{i}| + a_{i j}\cdot |\varepsilon_{i j}^{j}| \\
|\varepsilon_{i j}^{i}| + |\varepsilon_{i j}^{j}| \leq  v_{ij} \\
\text{NV}= \text{NV}^*\\
\text{constraints (\ref{cp33poip}-1)-(\ref{cp33poip}-8)}\\
  \end{array} \right.
  \end{array}
\end{equation}
where $|a_{i j} \exp(y_j) - \exp(y_i)| $ is a convex function, the other constraints are linear functions. Thus, this model is a mixed-integer convex programming model which can be solved by the nonlinear optimization solvers such as  \texttt{Ipopt}.

\subsection{Supplement for Section \ref{sec6}}

(1) \emph{The equivalent form of model (\ref{cp3poip11})}

Let $x_{ij}=\log \overline{{a}_{ij}}$, model (\ref{cp3poip11}) can be written as model (\ref{cp33poip12}) for all $\forall i, j, k, l \in N$:

\begin{equation}
\label{cp33poip12}
  \begin{array}{l}
  \min\quad J=\alpha \text{NPR}+\text{AOC}\\
  s.t.
    \left\{ \begin{array}{lll}
    -x_{ij}+x_{kl}- M(1 - l_{ijkl}^x)+ \varepsilon \leq  0,& (\ref{cp33poip12}-1)\\
    x_{ij}-x_{kl}- Ml_{ijkl}^x \leq0,  & (\ref{cp33poip12}-2)\\
     -(\overline{y_i}-\overline{y_j})+(\overline{y_k}-\overline{y_l})- \\M(1 - l_{ijkl}^x)+ \varepsilon \leq  0,  & (\ref{cp33poip12}-3)\\
    (\overline{y_i}-\overline{y_j})-(\overline{y_k}-\overline{y_l})- Ml_{ijkl}^x \leq0,  & (\ref{cp33poip12}-4)\\
    \sum\limits_{i = 1}^n \overline{y_i}=0  &(\ref{cp33poip12}-5)\\
    a_{ij}-\overline{a_{ij}} \leq M\delta_{ij}  & (\ref{cp33poip12}-6)\\
    -a_{ij}+\overline{a_{ij}} \leq M\delta_{ij}  & (\ref{cp33poip12}-7)\\

     \cal{D}(\mathbf{\overline{A}},\mathbf{\overline{W}})\leq \overline{\text{CI}},   &(\ref{cp33poip12}-8)\\
     x_{ij}+x_{ji}=0,   & (\ref{cp33poip12}-9)\\
    {x_{ij}} = \sum\limits_{m = 1}^{17}l_{ijm}c_m,  & (\ref{cp33poip12}-10)\\
\sum\limits_{m = 1}^{17}l_{ijm} = 1,   & (\ref{cp33poip12}-11)\\
\delta_{ij}, l_{ijkl}^x ,l_{ijm} \in {\rm{\{ }}0,1{\rm{\} }}  & (\ref{cp33poip12}-12)\\
  \end{array} \right.
  \end{array}
\end{equation}

In model (\ref{cp33poip12}), $l_{ijkl}^x$, $l_{ijm}$, $\delta_{ij}$, $\overline{y_i}$ and $x_{ij}$ are decision variables; $M$ is a big value, $\varepsilon$ is a positive number close to zero; $\mathbf{c}$ = $(-\log9,-\log8,-\log7,$ ${-\log6, -\log5, -\log4,-\log3,-\log2, 0, \log2, \log3,\log4,}$ ${\log 5,\log6, \log7,\log8,\log9)^T} $.

$l_{ijkl}^x$ is a  binary variable to represent the relationship between $x_{ij}$ and $x_{kl}$,
constraints  (\ref{cp33poip12}-1)-(\ref{cp33poip12}-2) are the necessary and sufficient conditions of $ l_{ijkl}^x=1 \Leftrightarrow x_{ij}>x_{kl}$, thus, constraints  (\ref{cp33poip12}-1)-(\ref{cp33poip12}-4) ensure that the priority vector of $\mathbf{\overline{A}}$ meets the POIP condition.
Constraints (\ref{cp33poip12}-6) -(\ref{cp33poip12}-7) are used to judge whether the original preference has been changed.
The modified PCM $\mathbf{\overline{A}}$  meeting the predetermined inconsistency level is designed as (\ref{cp33poip12}-8).
The  reciprocal property is modelled as (\ref{cp33poip12}-9).
$\overline{{a}_{ij}}\in DS_{[1/9,9]}$ is expressed as constraints (\ref{cp33poip12}-10) and (\ref{cp33poip12}-11),
 for example, if $l_{ij1}=1$, then $x_{ij}=c_1=-\log9$.
 
If we use the deviation function of model MEM in constraint (\ref{cp33poip12}-8), model (\ref{cp33poip12}) is a MIP model;
if we apply the  deviation function of model LLSM in constraint (\ref{cp33poip12}-8), model (\ref{cp33poip12}) is a MIQCP model;
if we apply the other deviation functions which are nonlinear but convex, model (\ref{cp33poip12}) is a MINP model.

The constraint (\ref{cp33poip12}-8) is a convex function in model (\ref{cp33poip12}), the other constraints are linear or mixed integer functions.
Hence, the optimal solution of model (\ref{cp33poip12}) can be easily obtained by using the existing mathematical programming optimizer such as Ipopt.
 We set $M=1,000$ and $\varepsilon=10^{-3}$, \texttt{Ipopt}  is applied to solve the above optimization models.

\end{document}